\documentclass[12pt]{article}
\usepackage{float}
\usepackage{amsmath}
\usepackage{amsfonts}
\usepackage{amssymb}
\usepackage{graphicx}
\usepackage{cite}
\usepackage{color}
\usepackage{dcolumn}
\usepackage{wrapfig}
\usepackage{longtable}
\usepackage{times}
\usepackage{caption}
\usepackage{subcaption}
\usepackage{bm}
\usepackage[margin=0.7in]{geometry}
\usepackage{newfloat}
\DeclareFloatingEnvironment[name={SI Figure}]{suppfigure}
\DeclareFloatingEnvironment[name={SI Table}]{supptable}

\begin{document}

\begin{flushleft}
{\LARGE
\textbf{Uncovering randomness and success in society}
}
\\
\vspace{0.3cm}
\bf Sarika Jalan$^{1,2,\ast}$,
Camellia Sarkar$^{2}$, 
Anagha Madhusudanan$^{1}$, 
Sanjiv Kumar Dwivedi$^{1}$
\\
\vspace{0.2cm}
\it ${^1}$ Complex Systems Lab, Physics Discipline, Indian Institute of Technology Indore, M-Block, IET-DAVV Campus, 
Khandwa Road, Indore 452017, India
\\
\it ${^2}$ Complex Systems Lab, Center for Biosciences and Biomedical Engineering, Indian Institute of Technology Indore, M-Block, IET-DAVV Campus, Khandwa Road, Indore 452017, India
\\
${^{\ast}}$ E-mail: sarika@iiti.ac.in
\end{flushleft}

\begin{abstract}
An understanding of how individuals shape and impact the evolution of society is vastly limited due to the unavailability of large-scale reliable datasets, that can simultaneously capture information regarding individual movements as well as social interactions. We believe that the popular Indian film industry, `Bollywood', can provide a social network apt for such a study. Bollywood provides massive amounts of real, unbiased data that spans over 100 years and hence this network has been used as a model for the present paper. It is seen that the nodes which maintain a moderate degree or widely cooperate with the other nodes of the network, tend to be more fit (measured as the success of the node in the industry) in comparison to the other nodes. The analysis carried forth in the current work, using a conjoined framework of complex network theory and random matrix theory, aims to quantify the elements that determine the fitness of an individual node and the factors that contribute to the robustness of a network. The authors of this paper believe that the method of study used in the current paper can be extended to study various other industries and organizations.
\end{abstract}

\section{Introduction}
The field of network analysis helps us to look at the study of an individual component as a part of a complex social structure and its interactions \cite{Barabasi_2002}. It explains various phenomena in a wide variety of disciplines ranging from physics to psychology to economics. The theory is adept at finding the causal relationships between network attributes such as the position of a node and the specific ties associated with it, and the fitness of the said node \cite{Borgatti_2009}.
Such relationships, that seemed thoroughly random to the eyes of a researcher only about a decade before, have now been vastly studied and documented \cite{Wasserman}. 
We aim to further investigate the very interesting idea that human behavior is predictable to a fair degree \cite{Barabasi_2010} using the Bollywood Network as a model for this purpose. 

Making nearly one thousand feature films and fifteen hundred short films per year, the Indian film industry is the largest in the world \cite{Focus_2010} which has held a large global population 
in more spheres of its existence than just entertainment. 
It mirrors a changing society capturing its peaks and valleys over time and impacts the opinions and views of the diverse populace \cite{Bose}. An example that can be stated as a proof of this was exhibited when the number of Indian tourists to Spain increased by $65\%$ in the year succeeding the box office success of the movie `Zindagi Na Milegi Dobara', which extensively portrayed tourist destinations in Spain, and also in the fact that Switzerland, depicted in various popular yesteryear Indian films (movies), remains a popular tourist destination for Indians to date \cite{FICCI_2011}.

The Hollywood co-actor network is a social network that has invited a fair amount of interest in the past \cite{Rev_Social_2006}, studies being conducted using relational dependency network analysis, Layered Label Propagation algorithm and 
PageRank algorithm \cite{Hollywood,Boldi_Vigna}. 
In comparison, its much larger counterpart in India has been largely ignored.
Flourishing with a $9\%$ growth from 2009 to 2010 \cite{FICCI_2011} and a 
further $11.5\%$ growth from 2010 to 2011 \cite{FICCI_2012}, it is an industry that sees blazingly fast growth, leading us to expect drastic changes in small time frames.
We study the Bollywood industry because it provides a fair ground to capture the temporal changes in a network owing to its rapidly changing character. Using data from the past 100 years, we construct a network for every five year period. The nodes can be classified into the three distinct categories : 1) lead male actors, 2) lead female actors and 3) supporting actors.
We analyze the structural properties of this network and further study its spectral properties using the random matrix theory (RMT). 

Though originally rooted in nuclear physics \cite{Mehta_1991}, RMT has found widespread applications in different real systems such as the stock-market indices, atmosphere, human EEG, large relay networks, biological networks and various other model networks. Under the framework of RMT, such systems and networks follow the universal Gaussian orthogonal ensemble (GOE) statistics. Though there exist other universality classes such as Gaussian unitary ensemble and Gaussian symplectic ensemble \cite{handbook_rmt}, which have also been
extensively investigated in RMT literature, we focus only on GOE statistics as spectra of various networks have been shown
to rest with this universality class \cite{SJ_2012,SJ_2007b,SJ_2009}.
The universality means that universal spectral behaviors, such as statistics of nearest neighbor spacing distribution (NNSD)
are not only confined to random matrices but get extended to other systems. A wide variety of complex systems fall under this
class, i.e. their spectra follow GOE statistics (\cite{Guhr} and references therein).

\section{Materials and Methods}
\subsection{Construction of Bollywood networks}
We collect all Bollywood data primarily from the movie repository website {\it www.bollywoodhungama.com} and henceforth from {\it www.imdb.com} and {\it www.fridayrelease.com} (now renamed as {\it www.bollywoodmdb.com}) and we generate no additional data.
The website {\it www.bollywoodhungama.com} previously known as {\it www.In
diafm.com}, is a reputed Bollywood entertainment website, owned by Hungama Digital Media Entertainment, which acquired Bollywood portal in 2000. We use Python code to extract names of all the movies and their corresponding information for a massive period of hundred years spanning from 1913 to 2012. Initially we document the names of all films as per their chronological sequence (latest to oldest) from the websites by incorporating the desired URL \cite{URL} in the code along with a built-in string function which takes the page numbers (932 pages in ``Released before 2012'' category and 24 pages in ``Released in 2012'' category) as input. Each film of every page bears a unique cast ID in the website, navigating to which via ``Movie Info'' provides us complete information about the film. In the Python code, we store the unique cast IDs of films in a temporary variable and retrieve relevant information using appropriate keywords from the respective html page. We also manually browse through other aforementioned websites in order to collect any yearwise missing data, if any. Thus we get the data in terms of names of the movies and names of the actors for 100 years. We then merge the data from all the websites and omit repetitions. A total of 8931 movies have been documented so far in Bollywood from 1913 till 2012. Harvesting the complete data took approximately 2000 hours of work over a 4-month period, which includes manual verification, formatting, removal of typos and compilation of the data. Considering the rapidly changing nature of the Bollywood network, we assort the curated massive Bollywood data in to 20 datasets each containing movie data for five-year window periods, as this is an apt time frame within which the network constructed is large enough to study the important network properties, and is not too large to miss any crucial evolutionary information. Since the number of movies and their actors in the time span 1913-1932 were scanty and could not have yielded any significant statistics, we merge the 1913-1932 datasets and present as a single dataset 1928-1932.

We create database of all actors who had appeared in the Bollywood film industry ever since its inception in five-year window periods,
as mentioned in the previous version of the manuscript, by extracting them from the movie information using Python algorithm and we assign a unique ID number to each actor in every span which we preserve throughout our analysis. We take care of ambiguities in spellings of names of actors presented in different websites by extensive thorough manual search and cross-checking to avoid overlapping of information and duplication of node identities while constructing networks.
Tracking by their unique ID numbers assigned by us, we create a co-actor database for each span where every pair of actors who had co-acted in a movie within those five years are documented. We then construct an adjacency list of all available combinations of co-actors. Treating every actor as a node and every co-actor association as a connection, we create a co-actor network of the largest connected component for every span.

We pick the actors appearing as the protagonist (occupant of the first position) in the movie star cast list from the movie star cast database created by us and observe that they incidentally are male actors in almost all movies with some rare exceptions. On extensive manual search based on popularity, award nominations we find that those male actors appear as a lead in the respective movies which made our attempt to extract lead male actors even easier. We could very well define the lead male actor as the protagonist in the star cast of at least five films in consecutive five-year spans and extract them from the movie star cast list using Python code while we were unable to find any proper definition for lead female actors as the second position of the movie star cast list is alternately occupied by either female actors or supporting actors, making it difficult to extract
them only based on the network data as described. Hence we handpick the lead female actors from the movie star cast database for all the spans based on their popularity, award nominations and create their database.

\subsection{Assimilation of Filmfare awards data}
We consider Filmfare award nominations as the best means to assess the success rates of all lead actors of Bollywood and distinguish the lead female actors from the rest. Filmfare awards were first introduced by the The Times Group \cite{filmfare} after the Central Board of Film Certification (CBFC) was founded by Indian central government in 1952 to secure the identity of Indian culture. The reason behind choosing Filmfare Awards amongst all other awards in our analysis is that it is voted both by the public and a committee of experts, thus gaining more acceptance over the years. Instead of the awards bagged we rather take into account the award nominations in order to avoid the interplay of some kind of bias affecting the decision of the CBFC committee in selecting the winner.
By manual navigation through every year of Filmfare awards available on the web, we create a database of all categories of Filmfare awards and extract their respective nominees chronologically from the html pages using Python codes. Henceforth we use C++ codes to count the number of times every actor is nominated in each five-year span. Thus we obtain a complete list of all actors in each span along with their number of Filmfare nominations.

\subsection{Structural attributes of Bollywood networks}
Considering $p_{k}$ to be the fraction of vertices with the degree k, the degree distribution of the constructed networks is plotted with $p_{k}$. It has been sufficiently proven that the degree distribution of real world networks are not random, most of them having a long right tail corresponding to values that are far above the mean \cite{Barabasi_2002}.

We define the betweenness centrality of a node i, as the fraction of shortest paths between node pairs that pass through the said node of interest \cite{Newman_2003}.
\begin{equation}
x_i=\sum_{st} \frac{n^i_{st}}{g_{st}}
\end{equation} 
where $n^i_{st}$ is the number of geodesic paths from $s$ to $t$ that passes through 
$i$ and $g_{st}$ is the total number of geodesic paths from $s$ to $t$.

\subsection{Measures used for success appraisal}
In the current work, the concept of a payoff has been borrowed from the field of management \cite{Jackson_1996}, and adapted to suit the Bollywood network analysis. Payoff has elucidated the success of the center and non-center agents in a unique efficient star 
network \cite{A_Watts_2001}. We use an improvised version of payoff as a means to assess success rates of the nodes in Bollywood.
For the purpose of devising net payoff ($P_i$), we study the datasets two at a time (accounting for ten years) and use the following definition:
\begin{equation}
P_i= \frac{1}{\Delta d_i} +  \langle \sin (\pi{d_n}) \rangle + \langle \sum_j w_j\left(\frac{1}{n_i}+\frac{1}{n_j}+\frac{1}{{n_i}{n_j}}\right) \rangle
\label{eq_net payoff}
\end{equation}
where, $\Delta d_i$ is the change in degree of a particular node $i$ in two consecutive spans. $d_n$ 
is its normalized degree in a particular span given as 
$d_n=(\frac{d_i - d_{min}}{d_{max} - d_{min}})$ with $d_i$ being the degree of the node i and $d_{max}$ and $d_{min}$ being the maximum and minimum degree in that particular span, respectively. The third term sums over all nodes j that node i has worked with where $n_i$ and $n_j$ are the number of movies that the node i and j has worked in respectively and $w_j$ the number of 
times the node $j$ has worked with the node $i$ in the considered time window. 
The averages denoted in the net payoff (Eq.~\ref{eq_net payoff}) refer to the values averaged over the two consecutive datasets.
Based on the values of $P_i$, the actors of every set studied were ranked and lists made.

Due to the absence of a unifying framework that can be used to evaluate the success of films and their actors in the years before the inception of Filmfare Awards in 1954, we restrict our analysis on assessment of success to the time periods spanning from 1954 and onwards. In order to adumbrate the success of actors in the industry, we define overlap as the intersection of sets of co-actors 
that an actor has worked with, in two consecutive time frames.

\subsection{Spectral analyses}
The random matrix studies of eigenvalue spectra consider two properties: (1) global
properties such as spectral distribution of eigenvalues $\rho(\lambda)$,
and (2) local properties such as eigenvalue fluctuations around $\rho(\lambda)$. Eigenvalue
fluctuations is the most popular one in RMT and is generally obtained from the
NNSD of eigenvalues.
We denote the eigenvalues of a network by $\lambda_i = 1,\hdots ,N$ and 
$\lambda_1>\lambda_2> \lambda_3> \hdots > \lambda_N$. 
In order to get universal properties of the 
fluctuations of eigenvalues, it is customary in RMT to unfold the eigenvalues 
by a transformation $\bar{\lambda_i}=  \bar{N}(\lambda_i)$, where $\bar{N}$ 
is average integrated eigenvalue density. Since we do not have any analytical 
form for $N$, we numerically unfold the spectrum by polynomial curve fitting 
\cite{Mehta_1991}. After 
unfolding, average spacings are unity, independent of the system. Using the 
unfolded spectra, spacings are calculated as $s^{(i)}= \bar{\lambda}_{i+1} - 
\bar{\lambda_{i}}$.

The NNSD is given by
\begin{equation}
%\label{first}
P(s) = \frac{\pi}{2}s\exp\left(-\frac{{\pi}s^2}{4}\right).
\label{eq_nnsd}
\end{equation}

For intermediate cases, the spacing distribution is described by Brody distribution as
\begin{equation}
P_{\beta}(s)=As^\beta\exp\left(-\alpha s^{\beta+1}\right)
\label{eq_brody}
\end{equation}
where $A$ and $\alpha$ are determined by the parameter $\beta$ as follows:
%\begin{equation}
\begin{align*}
A&=(1+\beta)\alpha, \, \, \, \alpha=\left[{\Gamma{\left(\frac{\beta+2}{\beta+1} \right)  }}\right]^{\beta+1}
\end{align*}
%\end{equation}
This is a semi-empirical formula characterized by parameter $\beta$. As $\beta$ goes from 0 to 1, the
Brody distribution smoothly changes from Poisson to GOE. Fitting spacing distributions of different
networks with the Brody distribution $P_{\beta}(s)$ gives an estimation of $\beta$, and
consequently identifies whether the spacing distribution of a given network is Poisson, GOE, or the
intermediate of the two \cite{Brody_1973}.

The NNSD accounts for the short range correlations in the eigenvalues. We probe for the long range correlations in eigenvalues using $\Delta_3(L)$ statistics
which measures the least-square deviation of the spectral 
staircase function representing average integrated eigenvalue density 
${N}(\bar\lambda)$ from the best fitted straight line for a finite interval of 
length $L$ of the spectrum and is given by
\begin{equation}
\label{eq_delta3}
\Delta_3 (L;x) = \frac{1}{L} min_{a,b}   \int_x^{x+L} [N(\overline{\lambda})-a\overline{\lambda}-b]^2 d\overline{\lambda}
\end{equation}
where $a$ and $b$ are regression coefficients obtained after least square fit. Average over several choices
 of x gives the spectral rigidity, the $\Delta_3(L)$. In case of GOE statistics, the $\Delta_3(L)$ depends logarithmically on L, i.e.
\begin{equation}
\Delta_3(L)  \backsim \frac{1}{\pi^2} \ln L
\label{eq_delta3_goe}
\end{equation}
\begin{figure}[h]
\centerline{\includegraphics[width=0.7\columnwidth]{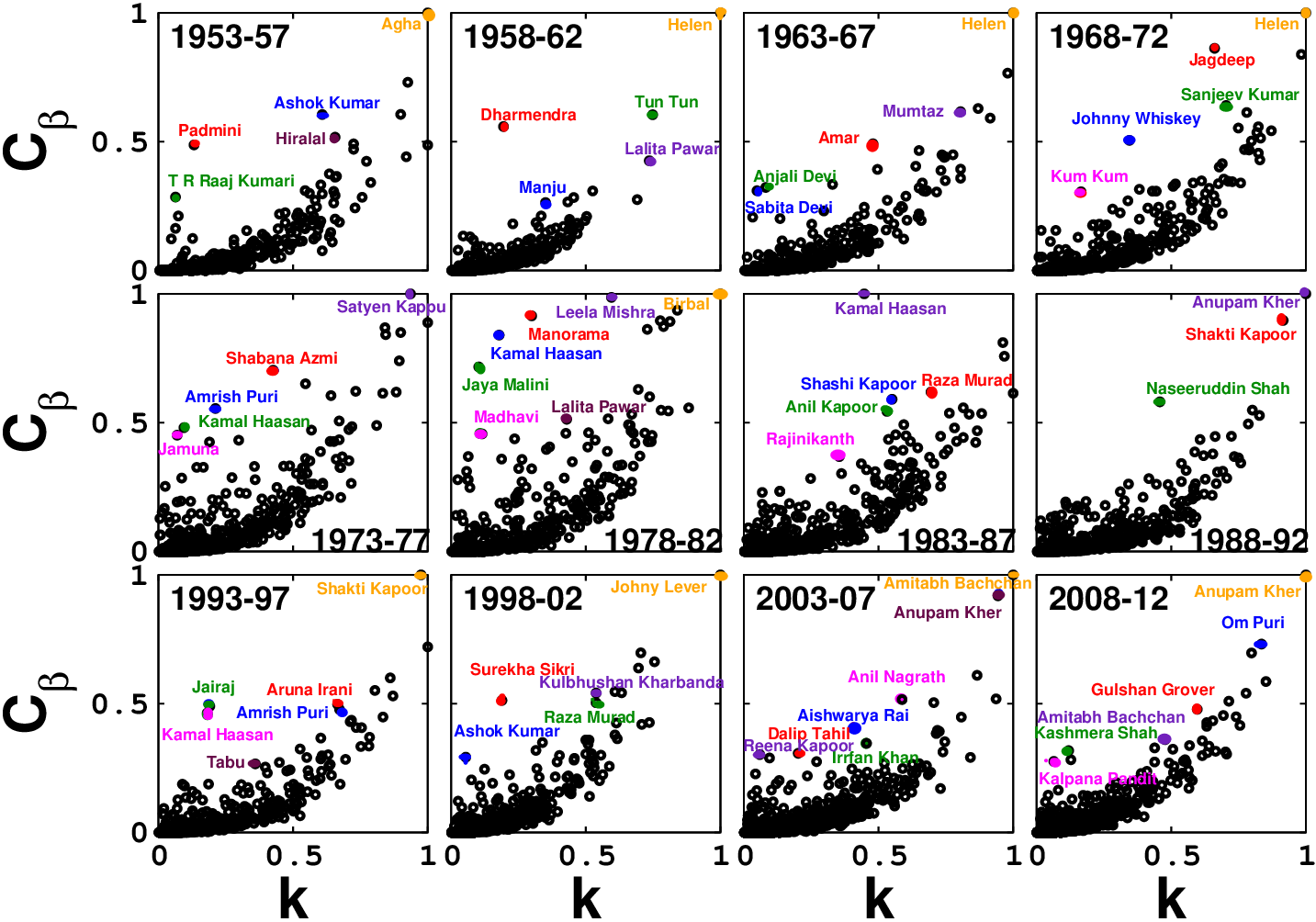}}
\caption{(Color online) Plots of normalized betweenness centrality ($C_{\beta}$) against normalized degrees ($k$) of 
Bollywood actors over 1953-2012. Actors and their corresponding betweenness centrality are represented in same color.  
}
\label{Figure1}
\vspace{-10pt}
\end{figure}
% Results and Discussion can be combined.
\section{Results and Discussion}
\subsection{Structural properties of Bollywood networks}
The degree distribution of the Bollywood networks follow power law, 
as expected based on the studies of other real world networks \cite{Barabasi_2002}. 
But an observation that defies 
intuition is that the most important nodes of the industry, acknowledged as the lead male actors, do not form the 
hubs of the constructed network, but instead have a moderate degree and also maintain it
along sets of data that were studied (SI Tables 1-6). Considering the network
on an evolutionary scale, this is a property that gains more prominence during the later sets of the data, 
while the network maintains power law throughout the entire timespan (SI Fig.~\ref{Deg_Dist}). 
The prominent supporting actors of the era form the hubs of the industry in respective time frames. This counterintuitive nature of the above observation can be explained by the fact that these actors collaborate with more nodes and take on more projects in a given time period. 
Hence they can be said to be instrumental in establishing connections in the network.
The scale-free behavior of the Bollywood industry can be elucidated by the fact that newcomers in the industry in general aspire to act with the lead actors of the era, who intuitively form associations with high degree nodes, thus illustrating the preferential attachment property prevalent in Bollywood networks \cite{Barabasi_2002}. 
\begin{figure}
\centerline{\includegraphics[width=0.6\columnwidth]{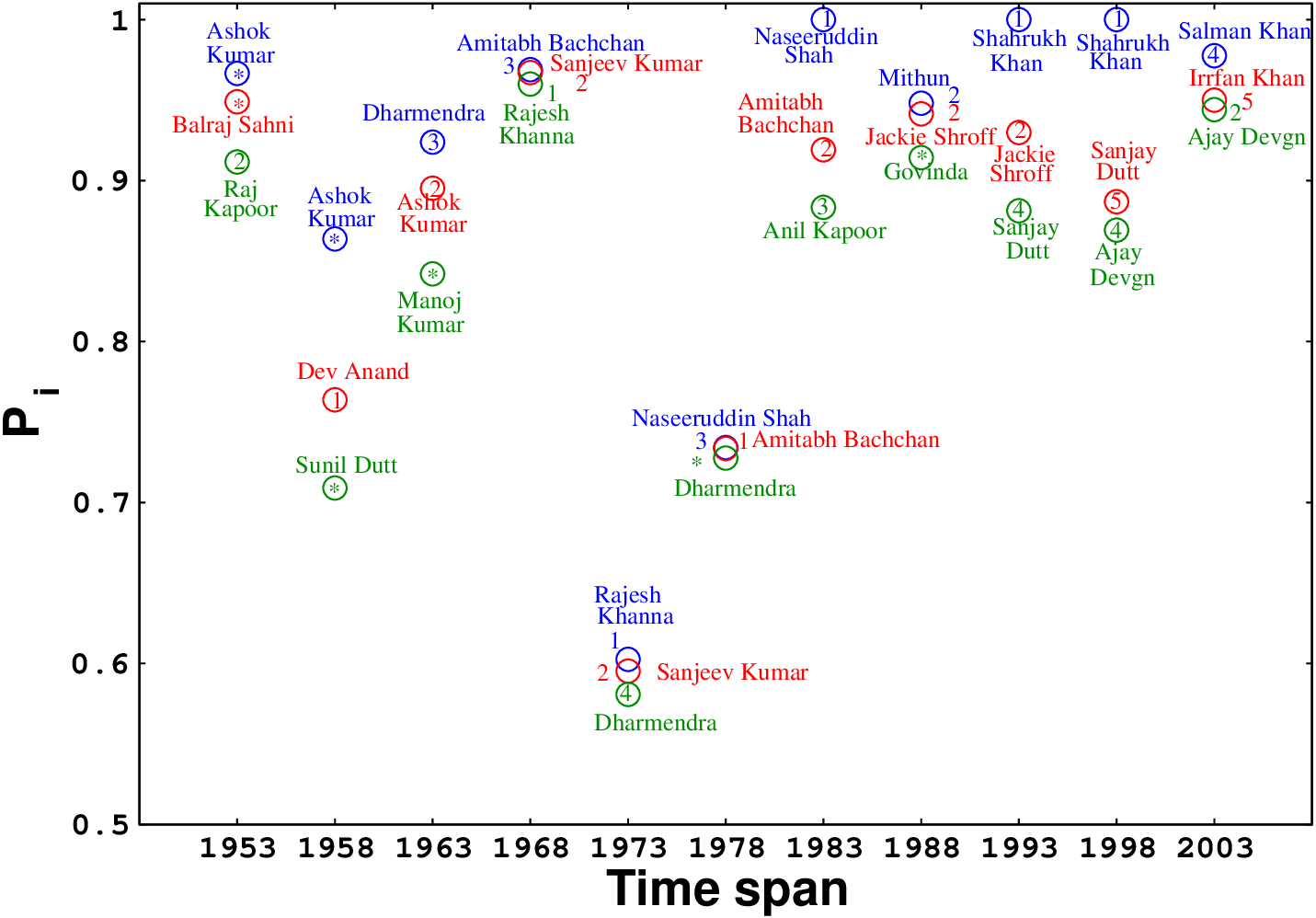}}
\caption{(Color online) Net payoff ($P_i$) of top three lead male actors. 
in each time span plotted against the respective time frames.   
They are ranked (as 1, 2 and so on) based on their number of Filmfare award nominations. `*' denotes no Filmfare award nominations. 
Actors and their corresponding rankings are represented in same color.
}
\label{Figure2}
\vspace{-10pt}
\end{figure}
\subsection{Success appraisal of Bollywood actors}
By virtue of the sinusoidal function used in (Eq.~\ref{eq_net payoff}), the nodes with a moderate degree
lead the net payoff list with both low degree and high degree nodes trailing behind. 
 The inverse of the change in degree favors nodes that preserve their degree over the years hence giving a higher net-payoff to actors who preserve their degrees over the various datasets. 

\begin{figure}
\centerline{\includegraphics[width=0.6\columnwidth]{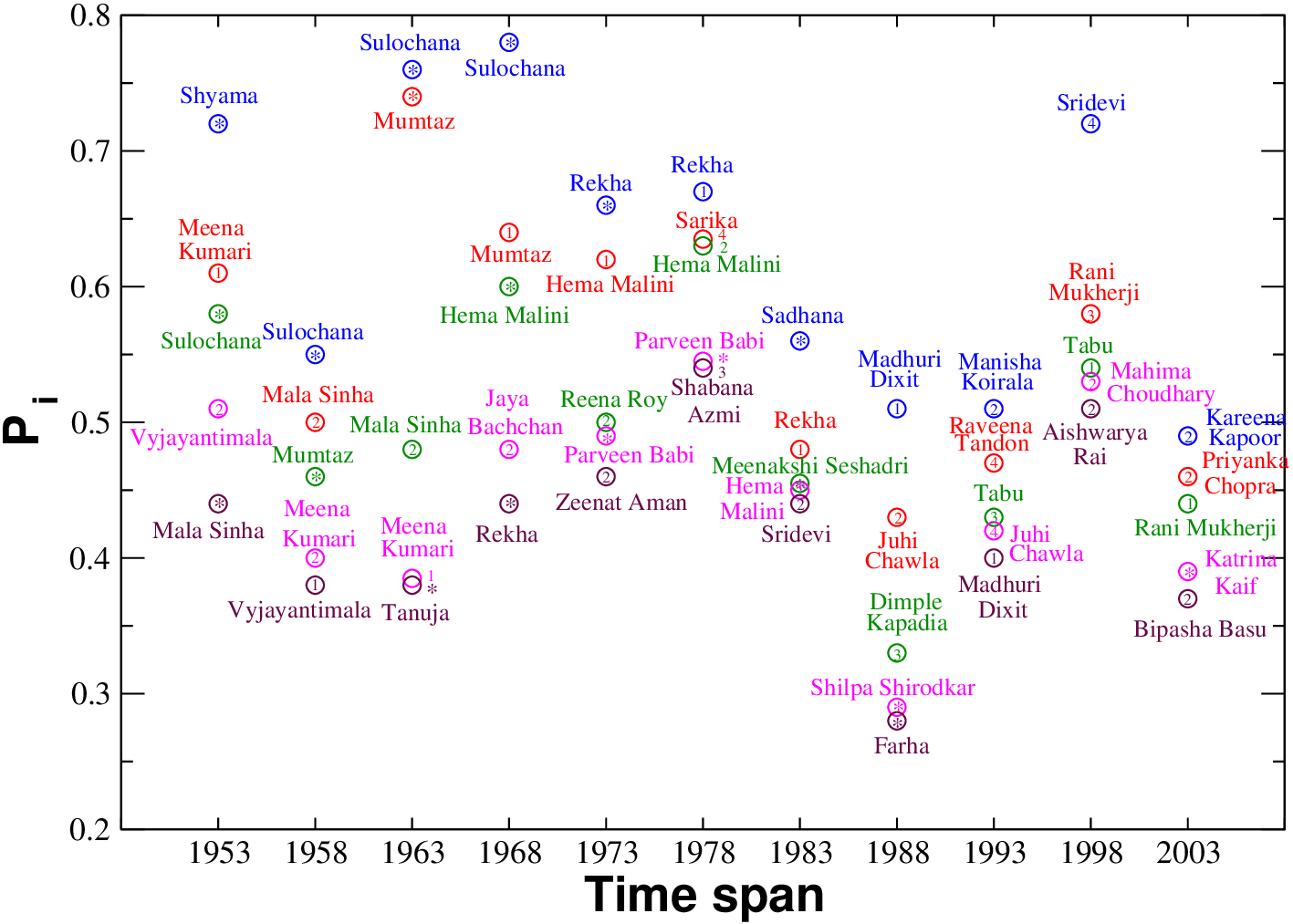}}
\caption{(Color online) Net payoff ($P_i$) of top five lead female actors 
in each time span plotted against the respective time frames. They are ranked (as 1, 2 and so on) based on their number of Filmfare award nominations. `*' denotes no Filmfare award nominations. Actors and their corresponding rankings are represented in same color.
}
\label{Figure3}
\vspace{-10pt}
\end{figure}

\begin{figure}
\centerline{\includegraphics[width=0.3\columnwidth]{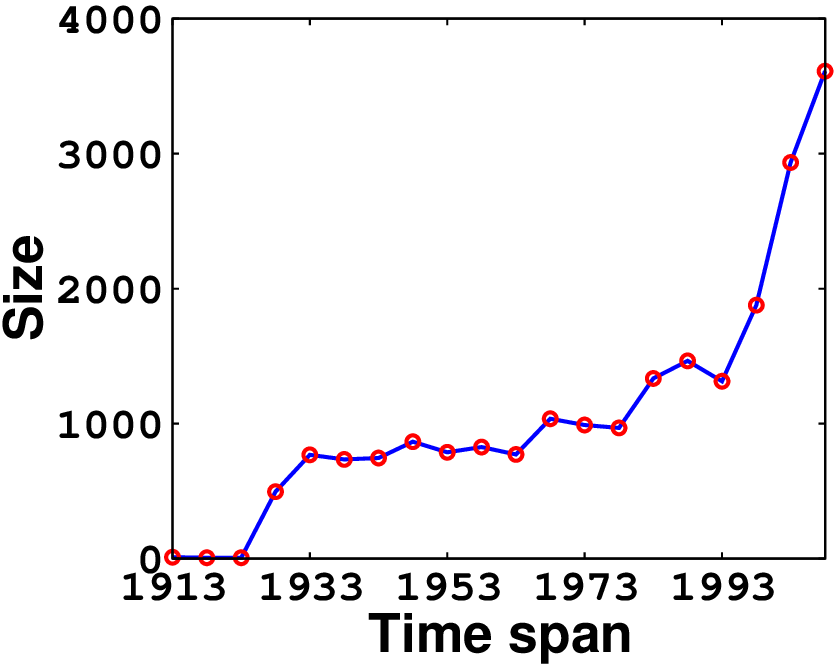}}
\caption{(Color online) Evolution of Bollywood network size over 1913-2012.
}
\label{Figure4}
\vspace{-10pt}
\end{figure}

\begin{figure}
\centerline{\includegraphics[width=0.6\columnwidth]{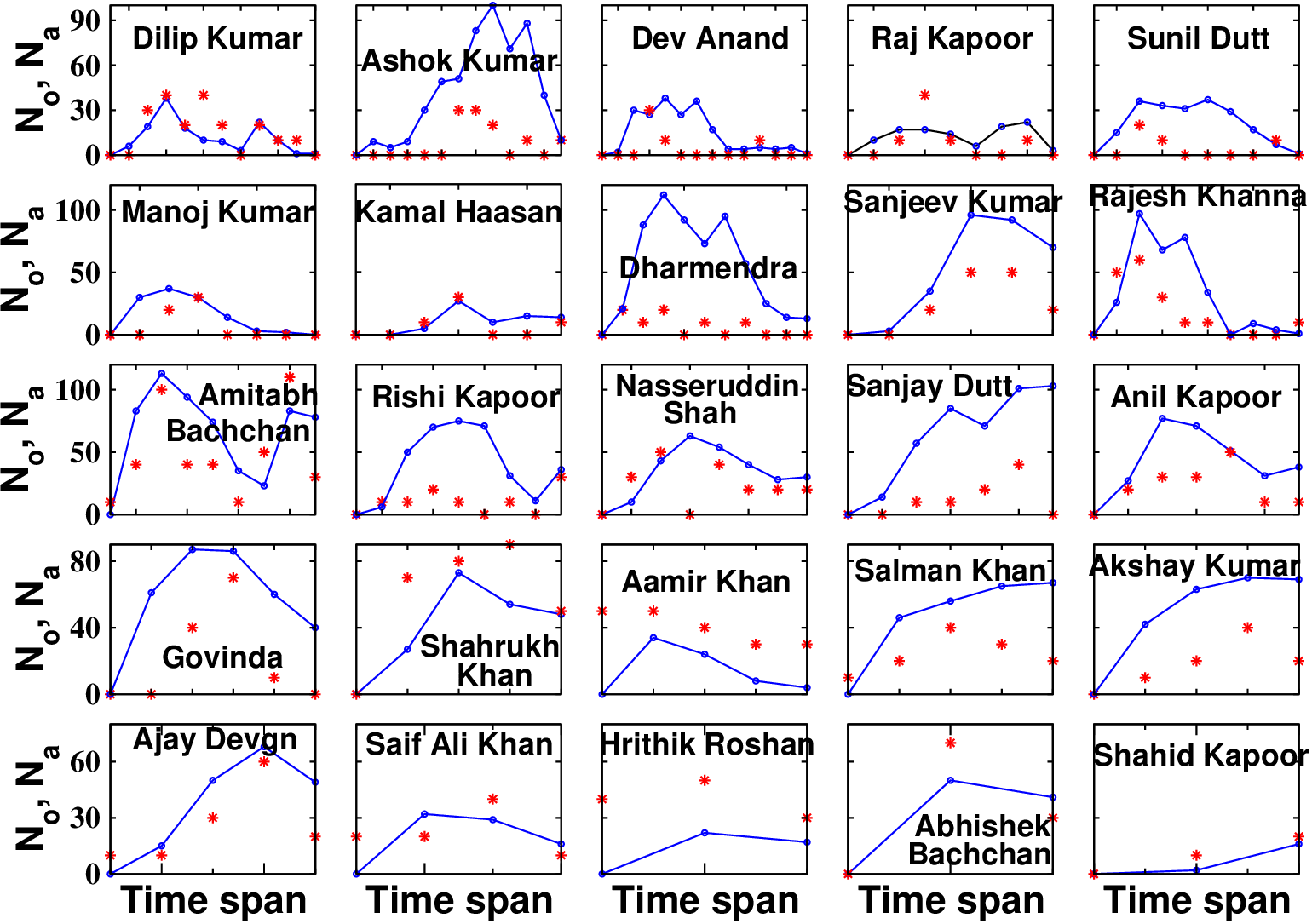}}
\caption{(Color online) Plots of individual overlaps $N_{o}$ (represented by $\bullet$)
of lead male actors and their Filmfare award nominations $N_{a}$ (represented by $\ast$) against their respective time spans. 
Time span here represents respective individual spans of lead male actors in Bollywood industry, for example Dilip 
Kumar had a long span stretching between 1943 and 1998 whereas Hrithik Roshan has a short spell 1998 onwards.  
}
\label{Figure5}
\vspace{-10pt}
\end{figure}

Successful supporting actors, although bear a high degree, appear quite high in the scale of $P_i$ because they have relatively higher values of $\langle p_i \rangle$.
Though interplay of various contrasting factors influence the appearance of lead male actors in $P_i$ list, 
they appear high in absolute scale of $P_i$ in all the sets under consideration except the ones corresponding to 1973-77 and 1978-82.
Three of the top five Filmfare award nominees in lead male actor category  appear 
as top three lead male actors in $P_i$ list in respective time frames (Fig.~\ref{Figure2} and SI Tables 1-6).
This observation is more pronounced in case of the lead female actors.
As observed in Fig.~\ref{Figure3} and SI Tables 7-12, the three lead female actors having secured the maximum 
number of Filmfare award nominations in a particular span of time, appear as the 
leading nodes in their respective $P_i$ list, a trait that is more consistent in the more recent datasets.
From the above analysis based on payoff it is supposed that possessing moderate degree and maintaining it are properties  
followed by the nodes that stand successful in Bollywood industry and can be contemplated as keys to success.    
Succeeding the economic liberalization in 1991, the inclusion of diverse socio-political-economic issues in mainstream Bollywood movies found favor with the audience \cite{task_force}. At around this period, Hollywood started gaining popularity among the Indian population owing to the advent of private movie channels and the internet. These factors coupled together affected the structure of the network, which might be the underlying reason behind the observed variations in the network properties, pre, post and during liberalization. A steep rise in the Bollywood network size 1993 onwards (Fig.~\ref{Figure4}) might
be one of the manifestations of this shift in economic policies. The status of an `industry' being conferred upon Bollywood in 1998 
might be a result of this increased size of the network \cite{Rita}. The comparatively larger shift of the network properties with the advent of liberalization as opposed to that caused by the introduction of the Filmfare awards in 1954, can lead us to conclude that mainstream Bollywood is largely driven by economic concerns rather than artistic ones.
The number of times an actor is nominated for the Filmfare awards while they remain a lead actor, when 
plotted with their overlap (as defined before), shows that 22 among the 25 actors exhibit an approximate direct proportionality (Fig.~\ref{Figure5})
emphasizing on the importance of winning combinations.
Overlap being one of the probable factors deciding the success of a node might explain the reason for the formation of 
social groups, and co-operation among them in the society \cite{Pacheco}.

High degree nodes indubitably have high betweenness centrality. Actors with high betweenness centrality seem to have a relatively larger
span in the industry even if their popularity levels, 
measured as the number of Filmfare award nominations, is not markedly high.
Nodes with the highest betweenness 
centrality of all datasets are found to be male actors (except Helen), whether lead or supporting, 
adumbrating the gender disparity in Bollywood.
Incidentally, few of the nodes bearing moderate and low degree also exhibit high betweenness centrality and also have a long span in the Bollywood industry (Fig.~\ref{Figure1}; SI Fig.~\ref{Between_deg} and Table 2).
This indicates that actors exhibiting mobility between diverse Bollywood circles seem to have an advantage of a long span,
though we are far from concluding that this is the only factor affecting the life span of a node. There exist examples
from social and biological systems which also support the importance of 
cooperation and mobility \cite{Lagoons}.
\subsection{Spectral analyses of Bollywood networks}
The spectral density, $\rho(\lambda$) of the connectivity matrix of Bollywood networks exhibit a triangular distribution 
(SI Fig.~\ref{Spectral_fig} and discussion in \cite{SI}), 
hence providing evidence supporting its scale-free nature \cite{Aguiar_2005}. The eigenvalue distribution of the Bollywood networks show a high degeneracy 
at $-1$, deviating from the commonly observed degeneracy at $0$ in most of the real world networks 
studied (for example, biological networks \cite{SJ_2012}) . 
This degeneracy at $-1$ can be attributed to the presence of clique structures in
the network \cite{book_spectra_graph}. 
Presence of dead-end vertices in spectrum and motif joining or duplication have been used as plausible explanations to widespread degeneracy at $0$ observed in biological networks \cite{Dorogovtsev}. 
\begin{figure}
\centerline{\includegraphics[width=0.6\columnwidth]{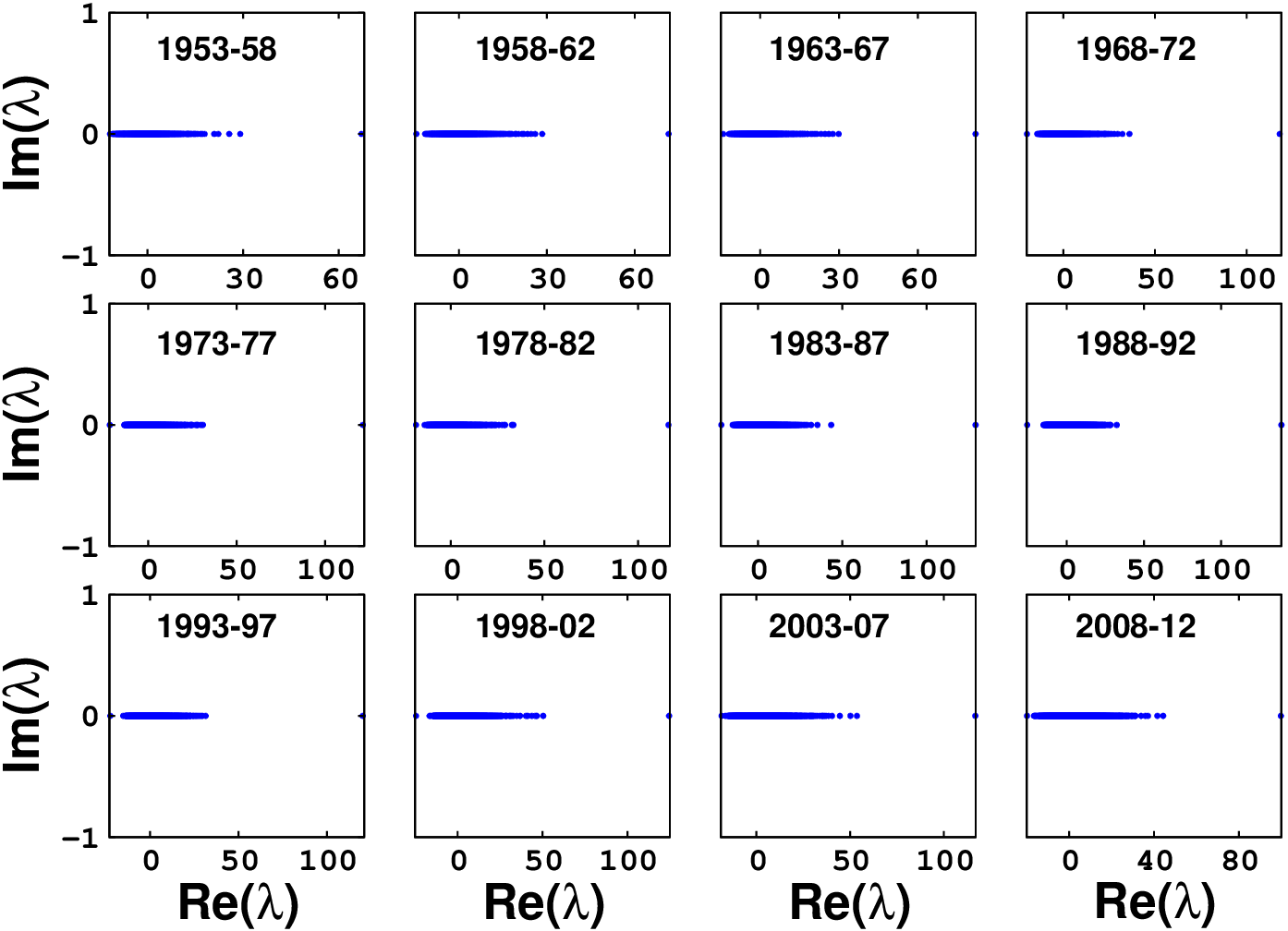}}
\caption{(Color online) Separation of lone eigenvalues from bulk of eigenvalues in Bollywood datasets spanning between 1953-2012.
}
\label{Figure6}
\vspace{-10pt}
\end{figure}
Factors affecting a social network are vastly different from those affecting a biological network, hence making the nature of their spectra varied. Owing to a relatively smaller number of nodes in the networks constructed for the periods 1913-17, 1918-22 and 1923-27, a bulk does not appear in their eigenvalue distributions. The distributions corresponding to the datasets of 1928-57, 1983-87 and 2003-12  very clearly show the presence of a few eigenvalues outside the bulk (SI Fig.~\ref{Bulk_Eigen} and Fig.~\ref{Figure6}), which is formed by the rest of the eigenvalues.  
While the largest eigenvalue is distinctly separated from the bulk, which is a well-known spectral feature of an
undirected network \cite{Newman_2003}, existence of other eigenvalues outside the bulk probably indicate the existence of 
distinct Bollywood guilds \cite{Chauhan_2009} further portending an evolving network structure. 

The spectral data as well as the data regarding the betweenness centrality of the networks, corresponding to the time periods after 1998-02, suggest that there has been a drastic change in the underlying network structure since then. This marked change in the more recent datasets in comparison to the older ones, is clearly illustrated by the presence of several eigenvalues outside the bulk (Fig.~\ref{Figure6}), and the presence of a lesser number of low degree nodes with a high betweenness centrality(Fig.~\ref{Figure1}). This indicates that the community structures in the Bollywood network have gotten more inter-interconnected post 1998-02, leading the authors of this paper to conclude that Bollywood is becoming increasingly systematic with time.

We fit the NNSD of Bollywood networks by the Brody distribution 
(Eq.~\ref{eq_brody}) and find that the value of 
$\beta$ comes out to be close to $1$ for all the datasets.
This implies that the NNSD of Bollywood datasets follow GOE statistics of RMT (Eq.~\ref{eq_nnsd} and SI Fig.~\ref{nnsd}) bringing Bollywood networks under the universality class of RMT \cite{Guhr,SJ_2007b}.
To examine the long range correlations, we calculate spectral rigidity via the $\Delta_3(L)$ statistics of RMT using Eq.~\ref{eq_delta3}
by taking same unfolded eigenvalues of different datasets as used for
the NNSD calculations. The value of $L$ for which the $\Delta_3(L)$ statistics follows RMT prediction (Eq.~\ref{eq_delta3_goe}) is given in the Table 1 and
the detailed plots are deferred to \cite{SI} as SI Fig.~\ref{Fig_Delta_3}. The $\Delta_3(L)$ statistics which provides a measure of randomness in 
networks \cite{SJ_2009} clearly indicate that the dataset corresponding to the 1963-67 timespan has the most random underlying network structure when compared with the other datasets. This notable feature of this timespan can probably be attributed to the consecutive wars that India was a part of in the years 1962 and 1965, which in turn lead to an extreme economic crisis in the country. As shown by the decreasing value of L since 1933, the networks have a trend of diminishing randomness.The dataset corresponding to 1948-52 witnessed a breach from this trend, probably due to the drastic political and financial changes post Indian Independence in 1947. One of the most crucial points exhibited in the analysis based on eigenvalue distribution and betweenness centrality is that, before the year 1998 the structure of the networks had either well segregated clusters or extreme random interactions, while post 1998 the structures seem to maintain a fairly consistent randomness (randomness measured by the value of L). 

\section{Conclusions}
Although Bollywood networks for different spans demonstrate varying amounts of randomness as suggested by the changing values of L in the $\Delta_3(L)$ statistics, observation of universal GOE statistics of the NNSD puts forward the evidence to show that a sufficient amount of randomness is possessed by all the sets. The efficiency of many real world systems such as the financial markets, the climatic system, neuronal systems etc, has been aided by their stochastic nature which leads to randomness \cite{nonlinear}. 
Bollywood network also provides an example to aid this relationship, as the industry has survived various valleys and crests since its inception, including in times of dire socio-economic crisis \cite{Economic_crisis}.
\begin{table}
\begin{center}
\caption{
\bf{Properties of Bollywood network of each 5 years block datasets.}
}
\begin{tabular}{|c|c|c|c|c|c|}      \hline
Time span & $N$	& $\langle k \rangle$  & ${N_eff}$ & $L$ & \%  ${\Delta_3}(L)$ \\ \hline
1928-32		& 	496	&   	9.46 		&	162	&	8	&	4.93			\\ \hline
1933-37		&	769	&	10.7 		&	246	&	6	&	2.43			\\ \hline
1938-42		&	735	&	13.3 		&	248	&	5	&	2.02			\\ \hline
1943-47		&	745	&	12.6 		&	276	&	5	&	1.81			\\ \hline
1948-52		&	866	&	17.5 		&	291	&	8	&	2.75			\\ \hline
1953-57		&	788	&	25.9 		&	272	& 	-	&	-			\\ \hline
1958-62		& 	827	&	29.9 		&	313	&      	-	&	-		\\ \hline
1963-67		& 	772	&	35.2 		&	308	&	19	&	6.16			\\ \hline
1968-72		& 	1036	&	47.0 		&	416	&	-	&	-		\\ \hline
1973-77		& 	990	&	47.5 		&	383	&	14	&	3.65			\\ \hline
1978-82		& 	968	&	45.1 		&	370	&	16	&	4.32			\\ \hline
1983-87		& 	1335	&	44.6 		&	480	&	19	&	3.95		\\ \hline
1988-92		& 	1465	&	44.9 		&	546	&	24	&	4.39		\\ \hline
1993-97		& 	1314	&	42.2 		&	504	&	12	&	2.38		\\ \hline
1998-02		& 	1878	&	46.3 		&	686	&	14	&	2.04		\\ \hline
2003-07		& 	2935	&	37.0 		&	973	&	17	&	1.74		\\ \hline
2008-12		& 	3611	&	30.3 		&	1164	&	17	&	1.46	\\ \hline
\end{tabular}
\end{center}
\begin{flushleft}$N$ and $\langle k \rangle$ respectively denote size and average
degree of network. $N_{eff}$ and $L$ are the effective dimension of non-degenerate eigenvalues less than
$-1$ and the length of the spectrum up to which spectra follow RMT. \%  The $\Delta_3(L)$ represents 
the extent of $L$ 2 which spectra follow GOE statistics, expressed in percentage terms. `-' denotes the
spectra which do not follow RMT.
\end{flushleft}
\end{table}
The extensive analyses of Bollywood data on the one hand reveals its influence on the decisions and preferences of the mass, while on the other it unravels the prevailing gender disparity \cite{gender_bias,book_gender} thus acting as a reflection of the society. Furthermore, it helps us deduce that cooperation among the nodes leads to combinations that become formulaic for successful ventures. It also seems to further propagate the idea suggesting that a combination of organization and randomness in the network structure supports the sustenance of the represented network. We believe that the analysis of the Bollywood network as carried out in this work can be extrapolated to study the predictability of success and the ingredients that are necessary for the robustness of other social collaboration networks
\cite{Social_collaboration} and organizations \cite{industries}.

% Do NOT remove this, even if you are not including acknowledgments
\section{Acknowledgments}
AM acknowledges IIT Indore for providing a conducive environment
for carrying out her internship. We are grateful to Arul Lakshminarayan (IITM)  
for time to time fruitful discussions on random matrix aspects and Dima Shepelyansky (Universit\'e Paul Sabatier) for useful suggestions. 
We are thankful to the Complex Systems Lab members, Ankit Agrawal and Aradhana Singh for helping with data download and
discussions.

\pagebreak

\begin{centering}
\underline{\LARGE{\textbf{Supporting Information}}}\\
\vspace{0.3cm}
\Large{\textbf{Uncovering randomness and success in society}}\\
\vspace{0.3cm}
\normalsize{\textbf{Sarika Jalan, Camellia Sarkar, Anagha Madhusudanan, Sanjiv Kumar Dwivedi}}\\
\end{centering}
\vspace{0.7cm}

\noindent {\large{\textbf{1. Methods}}} \\

\noindent Study of society and its movement has traditionally involved obtaining data from representative populations through field
studies and extrapolating the obtained results through approximations \cite{Newman_2001}. 
These methods of data collection provide, in the first place incomplete data and secondly, 
data that is prone to errors that would drastically skew the results obtained by the physicists' method of 
studying them. Movie actors networks analyses became a lucrative means for assessing society as the data obtained 
is to a satisfiable extent accurate and free from approximations and bias. \\

\noindent Although individual endowments (income) should rationally be the apt
discriminating factor for distinguishing lead actors from the supporting ones, it is quite cumbersome to retrieve relevant data due to
lack of reliable sources meant for the same. The variable nature of the data adds to its impediment.
We define lead male actors based on the number of times they top the starcast list in consecutive spans while defining lead female actors still remains an agony even after a century of cinematic heritage (discussed in sufficient detail in the main article). 
Although movies like Fashion, Page 3, Chandni Baar, Kahaani, Heroine portrays the never
ending struggle of women in society, the basis of their struggles have undoubtedly changed over the years. 
While Mother India (1957) depicts the struggle for existence, a struggle to combat
poverty, Fashion (2008) depicts a struggle for fame, a struggle for passion, a struggle for touching dreams, but not a struggle for
existence. This reflects a gradual change in the outlook of the society towards women. \\

\noindent In order to assess success of all actors in Bollywood industry, the Filmfare Awards were introduced for rewarding both artistic and technical excellence of
professionals in the Hindi language film industry of India. The National Film
Awards were also introduced in 1954 but gained less popularity as compared to Filmfare as they are decided by a panel appointed by Indian
Government and do not authentically reflect the choice of the global audience. The Filmfare Awards, in contrast, are voted for by both the
public and a committee of experts thus gaining more acceptance over the years. \\

\noindent {\large{\textbf{1.1 A brief review of Hollywood networks}}} \\

The collaboration graph of film actors were shown to be small-world networks \cite{Watts_1998} 
and their properties were studied using random graph theory \cite{Newman_2001b}.
Relational dependency network analysis has been performed on Hollywood datasets obtained from IMDB  
which identify and exploit cyclic relational dependencies to achieve significant performance gains \cite{RDN}.
Hollywood datasets were deployed for implementation of the Layered Label Propagation algorithm, 
meant to reorder very large graphs \cite{Boldi_2010} and the PageRank algorithm to uncover 
the relative importance of a node in a graph \cite{Vigna}. Professional links
between movie actors was used as a means to fit the predictions of a continuum theory to probe for the existence
of two regimes, the scale-free and the exponential regime \cite{Barabasi_2000}. \\
\vspace{1cm}

\noindent {\large{\textbf{1.2 Structural Analyses}}} \\

%\subsection{Structural Analyses}
\noindent {\textbf{1.2.1 Degree Distribution}} \\

%\subsubsection{Degree Distribution}
\begin{suppfigure}[!ht]
        \centering
        \begin{subfigure}[b]{0.4\textwidth}
                \centering
               \includegraphics[width=\columnwidth]{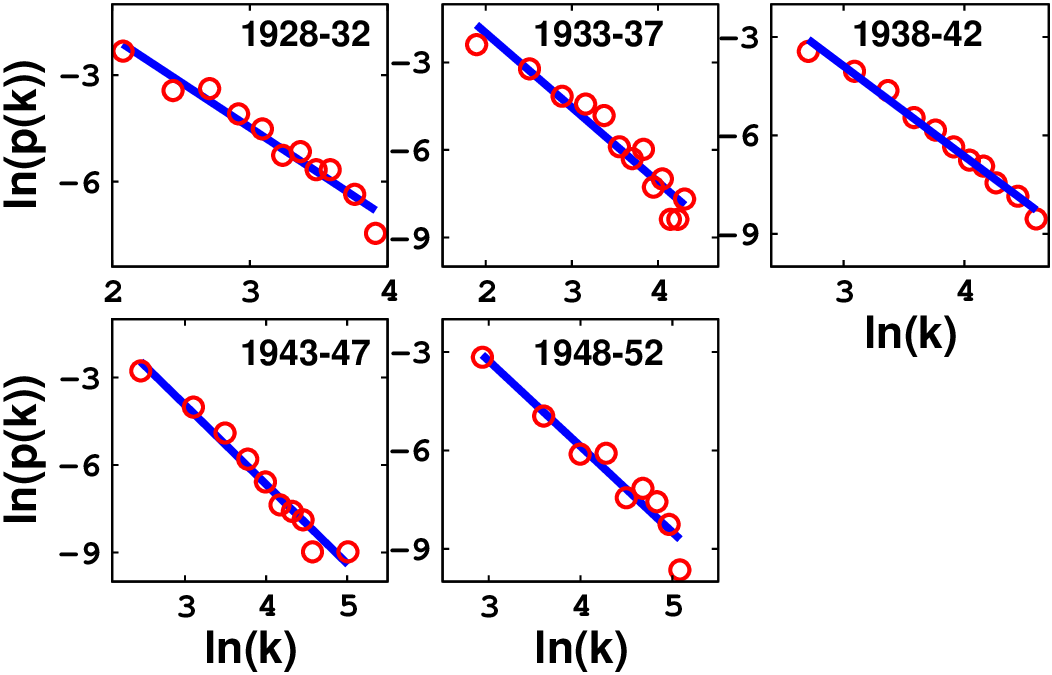}
        \end{subfigure}
        \quad 
        \begin{subfigure}[b]{0.5\textwidth}
                \centering
              \includegraphics[width=0.9\columnwidth]{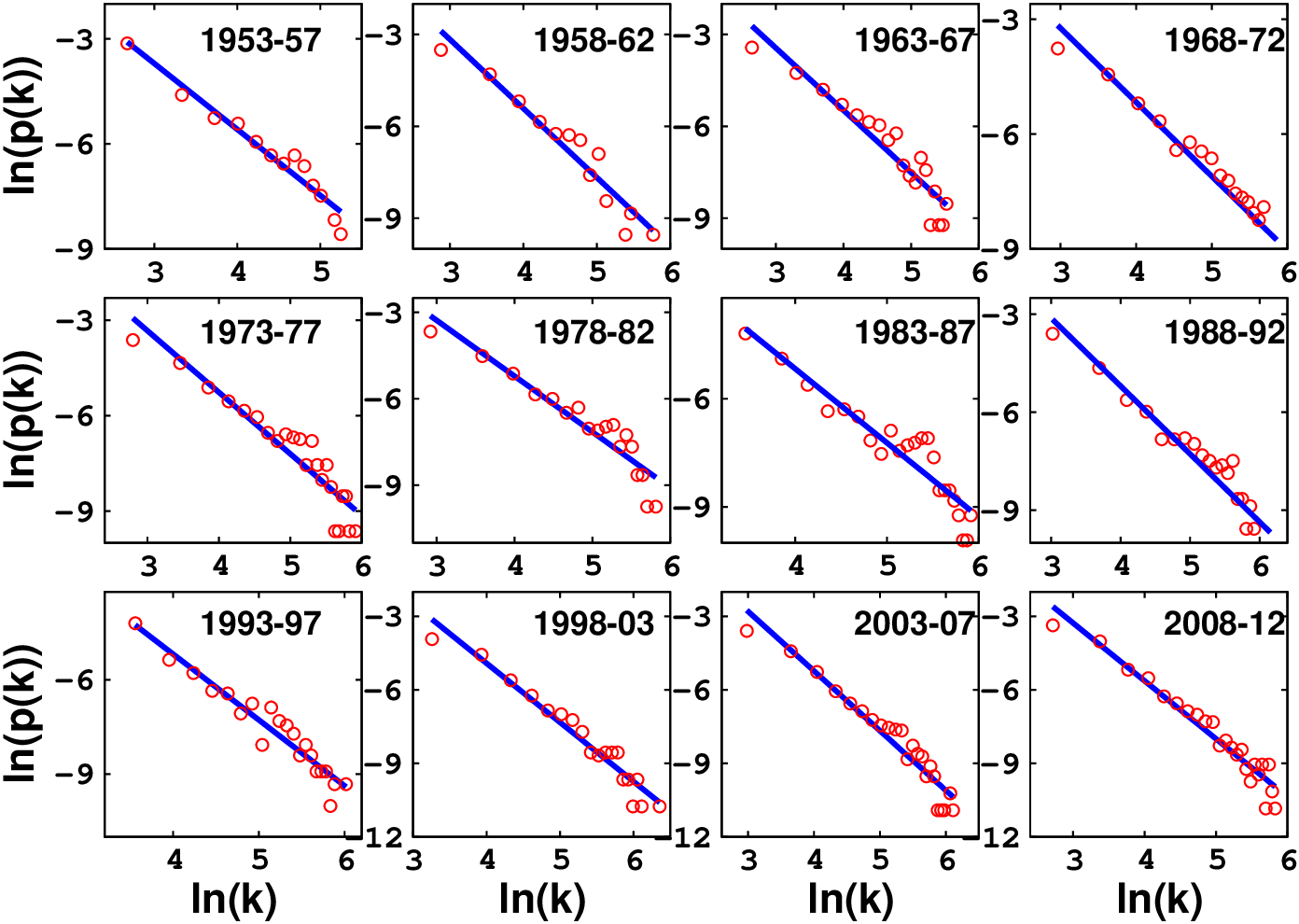}
        \end{subfigure}
        \caption{Degree distribution of the Bollywood networks over 1913-2012. Due to scarcity of actors in 1913-1927, all nodes appearing in 1913-27 have been merged and included in 1928-32.}
	\label{Deg_Dist}
\end{suppfigure}
\noindent Degree of a node can be defined as the number of nodes that are linked to the said node. 
Degree distribution is the plot of the degree versus the number of nodes with the particular degree. SI Fig.~\ref{Deg_Dist} plots degree distribution of Bollywood networks.\\
 
\noindent {\textbf{1.2.2 Betweenness Centrality}} \\

%\subsubsection{Betweenness Centrality}
\noindent The supporting actors have been observed to have high betweenness centrality. Nodes having higher degree would naturally be
coming into shortest path between pair of nodes, and hence would have high betweenness centrality. 
Fig.~\ref{Figure4} of main article and SI Fig.\ref{Between_deg} has highest $C_\beta$ corresponding to node possessing highest degree.
The fact that larger degree in any of the sets in 1928-2012 are possessed by supporting actors, and it is somewhat established that
supporting actors have longer life span
than lead male actor and lead female actors, makes the positive correlation between degree and life span quite
obvious. 
\begin{suppfigure}[h!]
\begin{center}
\includegraphics[width=0.2\columnwidth]{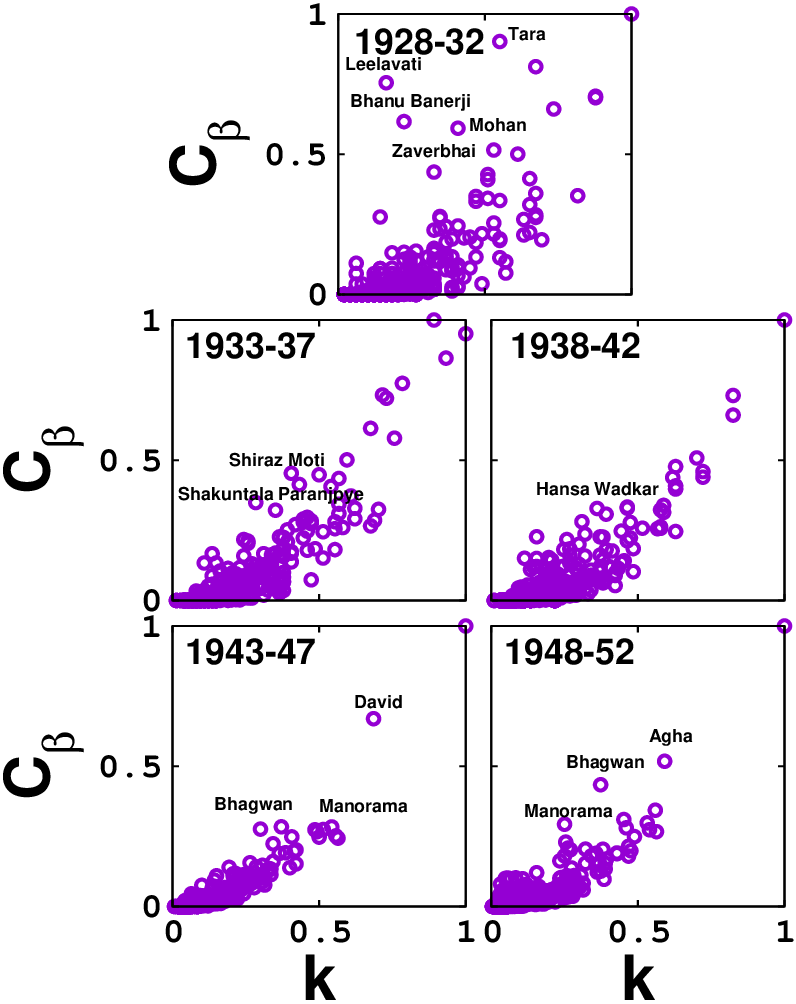}
\caption{Plots of normalized betweenness centrality ($C_{\beta}$) against normalized degrees ($k_{}$) of Bollywood actors over 1913-1952. }
\label{Between_deg}
\vspace{-20pt}
\end{center}
\end{suppfigure}
But some of the low degree nodes are also seen to have high betweenness centrality. Either they are supporting actors which again comply with the
earlier argument for their larger life span, or if they are lead male actors then also they show accredited life span. 
For example, in 1958-62 dataset, Dharmendra having low degree
distinctly appears in the high betweenness centrality region and has a remarkably long span (1953-2012) in the industry. Few other
prominent actors who have been seen to follow this trend are Kamal Haasan (1958-2012), Nasseruddin Shah (1973-2012), 
Rajinikanth (1973-2012), Anil Kapoor (1978-2012). These examples are taken for
those who are clearly depicting high betweenness centrality than rest of the nodes around them. 
Various female actors having low degree also fall in high betweenness centrality region and have long span. 
Padmini (1948-77) and Rajinikanth (1973-2012) are 
Tamil actors who have been observed in high betweenness centrality region bridging the gap between communities of Bollywood and Kollywood (Table 2). \\

%\begin{supptable}
\begin{center}
\begin{longtable}{|p{2 cm}|p{2cm}|p{10 cm}|}	
\caption{List of prominent actors who appear high in betweenness centrality zone} \\ \hline
\multicolumn{1}{|p{2cm}|}{Names of actors} & \multicolumn{1}{|p{2cm}|} {Span} & \multicolumn{1}{|p{10cm}|}{Recognition}  \\ \hline
\endfirsthead
\multicolumn{3}{c}
{{\tablename\ \thetable{} --- continued }}\\
\endhead

\hline  \multicolumn{3}{|r|} {{continued }}\\ \hline 
\endfoot
\hline \hline
\endlastfoot
Agha	& 1937-1989		& Known for comic roles, won Filmfare Best Supporting Actor Award (1960)     \\  \hline 
Ashok Kumar	&	1936-1993  & An iconic figure in Indian cinema popularly known as ``Dadamoni" 
who is also a painter, homeopath, astrologer, boxer, 
chess player, singer ; confered with honors like 
Dadasaheb Phalke award (1988) and Padma Bhushan (1998), Filmfare Lifetime Achievement Award (1995), Sangeet Natak Akademi Award (1959), National Film Awards for Best Actor (1969), Filmfare awards (1962, 1966, 1969)  	\\  \hline 
Padmini	&	1948-1994 &  An elegant {\bf Tamil} dancer who was also featured in several Hindi films; won Filmfare Award for Best Supporting Actress (1966)	\\  \hline 
Hiralal	&	1928-1995	& A prominent supporting actor having a long span in industry	\\  \hline 
T R Rajakumari	&	1936-1955  & Originally a Tamil film actress, Carnatic singer and dancer also acted in many Bollywood films \\    
\hline 
Helen	&	1951-2012	& An Indian film actress and one of the most popular dancers of all times; has bagged Padma Shri (2009), Filmfare Best Supporting Actress Award (1979), Filmfare Lifetime Achievement Award (1998)	\\  \hline 
Tun Tun	&	1946-1990	& A highly rated playback singer who later became a permanent comic relief in numerous 
Bollywood films. 	\\  \hline 
Dharmendra	&	1960-2012  & Often referred to as the ``He-Man", he has won Padma Bhushan (2012), Filmfare Lifetime 
Achievement award (1997), Filmfare Best Actor awards (1967, 1972, 1974, 1975), the Living Legend award (FICCI) and many more	\\  \hline 
Lalita Pawar	&	1928-1997  & Known for her roles as wicked matriarch and mother-in-law, she has won Filmfare Best 
Supporting Actress Award (1959) and Sangeet Natak Akademi Award (1961)	\\  \hline 
Mumtaz	&	1952-1976	& Critically acclaimed highly paid actress who has bagged a Filmfare Award for Best Actress 
(1970) and Filmfare Lifetime Achievement Award (1996) 	\\  \hline 
Anjali Devi	&	1936-1994  & A veteran Telugu and Tamil actress well known for her mythological roles in Bollywood \\  \hline 
Sabita Devi	&	1924-1996   & Supporting female actor	\\  \hline
Jagdeep	&	1951-2012  & Especially known for his excellent comic timing and appearances in horror movies and character 
roles. \\  \hline  
Sanjeev Kumar  &	1960-1985 & An accomplished Indian film actor remembered for his versatility and genuine portrayals of 
characters; has bagged National Film Award for Best Actor (1971, 1973), Filmfare Award for Best Actor (1976, 1977)	\\  \hline 
Johnny Whisky	&	1961-1997	& Popular supporting male actor	\\  \hline 
Kum Kum	&	1954-1973  & With her sumptuous dancing talent, she has starred with superstars of the era	\\  \hline 
Satyen Kappu	&	1952-2007  & A remembered character actor of Bollywood films	\\  \hline 
Shabana Azmi	&	1974-2013	& Regarded as one of the finest Indian actress of film, television and theatre proficient in 
a variety of genres with a record of five wins of the National Film Award for Best Actress (1975, 1983, 1984, 1985, 1999), 
Filmfare Best Actress award (1978, 1984, 1985), Filmfare Lifetime Achievement award (2006) and several international honours \\  \hline 
Amrish Puri	&	1954-2005	& Primarily remembered for essaying iconic negative roles in Bollywood and international film 
industries; has Filmfare Best Supporting Actor awards (1986, 1997, 1998), Sangeet Natak Akademi Award (1979)   \\  \hline 
Kamal Haasan	&	1959-2013  & Critically acclaimed Indian film actor, screenwriter, producer, director, songwriter, playback singer
and choreographer; has won a record 19 Filmfare Awards ranging across five languages, four National Film Awards,  Padma Shri, one
Rashtrapati Award for Best Child Artist and several other state, national and international honours.	\\  \hline 
Jamuna	&	1954-1968  & A veteran Telugu actress who has also won Filmfare Best Supporting Actress award (1968) for a 
Hindi movie. \\  \hline 
Birbal	&	1966-2011	& A veteran comedian who has acted in 377 Bollywood films.	\\  \hline 
Leela Mishra	&	1936-1986  & A character actress with roles varying from mothers, benign or evil aunt to comic roles; 
has acted in over 200 Hindi films	\\  \hline 
Manorama	&	1941-2005	& A Bollywood character actress, acted in over 160 films, known best for her role as the 
comical tyrant mother or villainous roles	\\  \hline 
Jaya Malini	&	1976-1988	& Has acted in over five different languages; known for her dance and vamp roles 	\\  \hline 
Madhavi	&	1981-1994  & Indian film actress acted in 7 languages in about 300 films	\\  \hline 
Raza Murad	&	1965-2013  & With a rich baritone voice, he often portrays negative character roles	\\  \hline 
Shashi Kapoor	&	1941-1999	&  An award-winning Indian film actor, director and producerPadma Bhushan	\\  \hline 
Anil Kapoor	&	1980-2013	& One of the most successful actors of Bollywood with National Film Award for Best Actor (2001), 
Feature Film (2008), Filmfare Best Actor Award (1989, 1993, 98), Filmfare Best Supporting Actor Award (1985, 2000)	\\  \hline 
Rajinikanth	&	1975-2013 & Being one of the highest paid actors of Asia, he is a cultural icon holding a matinee idol status; 
has been bestowed Padma Bhushan (2000)	\\  \hline 
Anupam Kher	&	1982-2013  & A versatile Indian actor who has appeared in nearly 450 films and 100 plays in almost all 
possible genres including international Oscar nominated films; honoured with Padma Shri (2004), National Film awards (1989, 2005),
Filmfare awards (1984, 1988, 1989, 1990, 1991, 1992, 1993, 1995)	\\  \hline 
Shakti Kapoor	&	1978-2012   & One of the leading villains in Bollywood movies also applauded for his comic roles; 
bagged Filmfare Best Comedian Award (1995)	\\  \hline
Naseeruddin Shah	&	1972-2013  & Considered to be one of the finest Indian stage and film actors; recipient of Padma Shri
(1987), Padma Bhushan (2003), National Film awards (1979, 1984, 2006), Filmfare awards (1981, 1982, 1984, 1993, 1995, 1996, 1998, 2000,
2007, 2008), Best Actor Venice Film Festival (1984)	\\  \hline  
Aruna Irani  &	1961-2010  & A popular supporting actress, has acted in over 300 films Filmfare Best Supporting Actress Award (1985, 1993), Filmfare Lifetime Achievement Award (2012)	\\  \hline 
Jairaj	&	1929-1995	& A renowned film actor, director and producer; recipient of Dadasaheb Phalke Award for lifetime
achievement (1980)	\\  \hline 
Tabu	&	1980-2013  & Garnered critical appreciation for acting in artistic, low-budget films across five languages; won Padma Shri
(2011), National Film Award for Best Actress (1997, 2002), Filmfare awards (1995, 1998, 2000, 2001, 2007)	\\  \hline 
Johny Lever	&	1984-2013  & One of the most popular comedians in Hindi cinema, has won Filmfare Best Comedian Award (1998, 1999)
including 13 nominations, \\  \hline 
Kulbhushan Kharbanda	&	1974-2013  & A popular Indian film, television actor, has been portrayed in a variety of roles ranging from
a bald villian, doctor, police, hero to character roles; nominated for Filmfare Best Supporting Actor Award (1986)	\\  \hline 
Surekha Sikri	& 1978-2006	  & An Indian film, theatre and TV actress recently popular as the negative diva of telly wood, has won
National Film Award for Best Supporting Actress (1988, 1995), Sangeet Natak Akademi Award (1989)	\\  \hline 
Anil Nagrath	&	1966-2013  & Popular supporting actor	\\  \hline 
Aishwarya Rai	&	1997-2013	& Winner of Miss India and Miss World pageants (1994) is a leading contemporary actress of Indian cinema
proficient in a range of genres; Padma Shri (2009), Filmfare Best Actress Award (1999, 2002), Most Glamorous Star of the Year (2007),
Outstanding Achievement in International Cinema (2009), Decade of Global Achievement Honour (FICCI, 2011)	\\  \hline 
Dalip Tahil	&	1974-2012	& Indian film, television and theatre actor known primarily for his negative roles has also demonstrated
his versatality playing character roles in a series of national and international television serials and films 	\\  \hline 
Irrfan Khan	&	1988-2013 & India's best known international actor skilled in performing in a variety of genres; has Padma Shri
(2011), Filmfare Awards (2003, 2007, 2012), Screen Actors Guild Award (2008), IRDS Film Award for social concern (2012) to his credit        
\\  \hline 
Gulshan Grover	&	1980-2013  & An Indian actor and film producer known for his villainous roles and later for his comic 
roles as well; has many national and international honours to his credit	\\  \hline 
Kashmera Shah	&	1994-2011	& An Indian actress and model who has won beauty contests	\\  \hline 
Om Puri	&	1976-2013	& Critically acclaimed for his performances in many unconventional roles in both mainstream Indian films and 
art films; winner of Padmashri (1990),  National Film Award for Best Actor (1982, 1984), Filmfare awards (1981, 2009), 
Karlovy Vary International Film Festival Best Actor (1984), Brussels International Film Festival Best Actor (1998), 
Grand Prix Special des Amériques Montréal World Film Festival for cinematographic art (1998)	\\  \hline 
Kalpana Pandit	&	2000-2013  & An emergency physician, who turned into an Indian film actress and model; 
has hosted technical awards ceremony and has made red carpet appearances at Hollywood premier nights	\\  \hline 
Reena Kapoor	&	2000-2013  & An Indian actress in films and television serials.	\\  \hline 
\end{longtable}
\label{betweenness_table}
\end{center}
%\end{supptable}

\noindent {\large{\textbf{1.3 Spectral Analyses}}} \\

%\subsection{Spectral Analyses}
\noindent Paul Erd\"os and Alfred R\'enyi pioneered the study of random graph models \cite{random_graph}, 
which persisted as a preferred method for
studying networks for decades. Following this, the Barab\'asi-Albert model of networks suggested that many
complex networks follow a power  
law degree distribution, hence forming what is termed as scale free network, which emerged as a revolutionizing change in network  
analysis and completely changed the perspectives of the analysts \cite{Barabasi_1999}. Some of the popular networks studied henceforth 
namely the Internet, the World-Wide-Web, cellular networks, phone call networks, science collaboration networks etc. appeared to follow 
the power law distribution \cite{Barabasi_2002}. 
For the undirected networks constructed here all the eigenvalues are real. We observe a high degeneracy 
at $\lambda = -1$, with almost 40\%
of states having this value. The presence of degeneracy at -1 is attributed to abundance of clique structure in underlying network
probably arising due to several actors appearing in a same movie. Eigenvalue statistics of Bollywood network elucidate typical triangular
structure, as observed for scale free networks \cite{Farkas_2003, Dorogovstev_2003}, with a crucial difference in 
having peak at -1 (SI Fig.~\ref{Spectral_fig}). \\

\begin{suppfigure}[h!]
\begin{center}
\includegraphics[width=0.5\columnwidth]{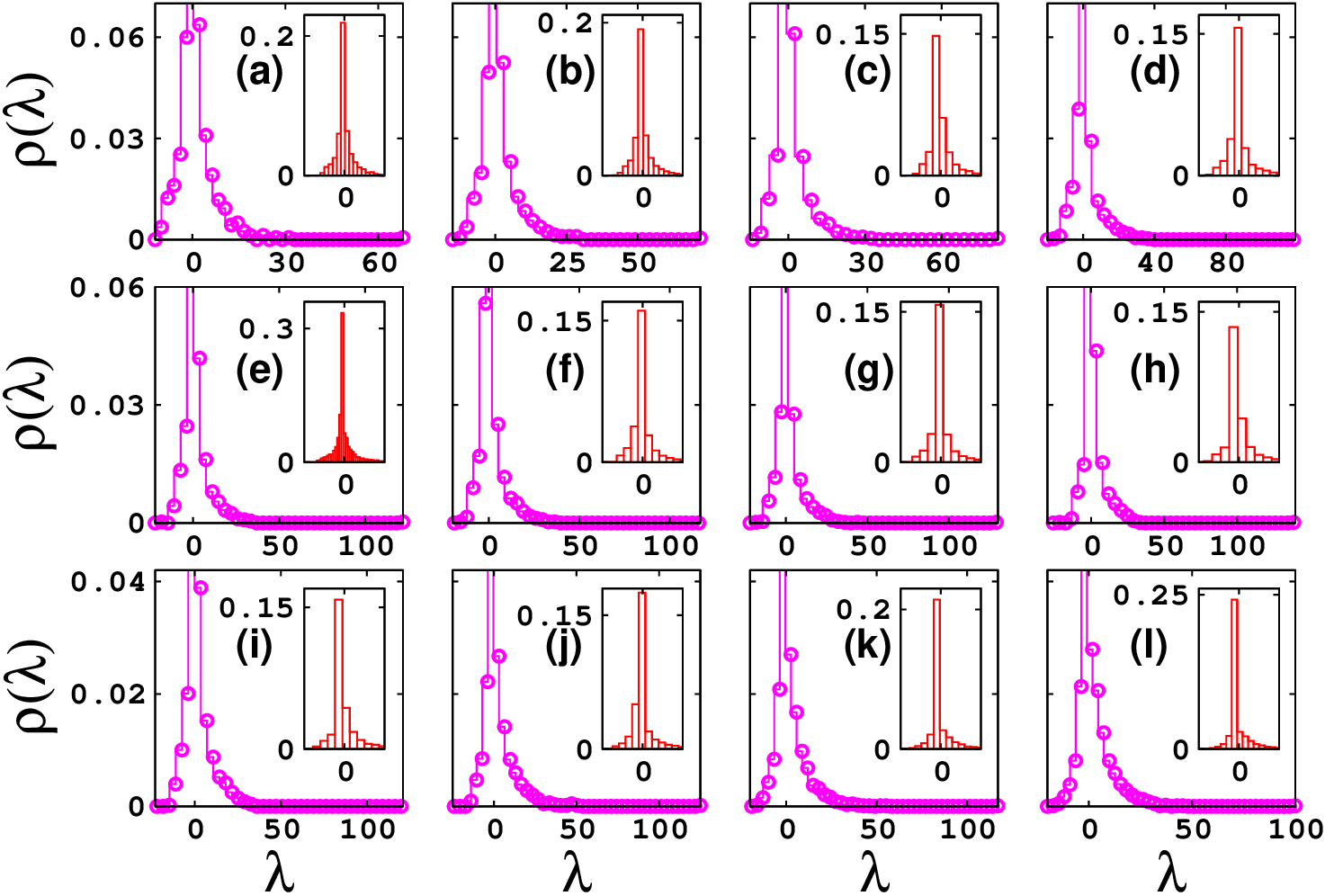}
\end{center}
\vspace{-20pt}
\caption{Spectral density distribution $\rho(\lambda$) of Bollywood networks. [(a)-(l) stand for 1953-57,
1958-62, 1963-67, 1968-72, 1973-77, 1978-82, 1983-87, 1988-92, 1993-97, 1998-02,
2003-07 and 2008-12, respectively]. Inset depicts peak of distribution.}
\vspace{-10pt}
\label{Spectral_fig}
\end{suppfigure}

\noindent Eigenvalue plots of Bollywood datasets (SI Fig.~\ref{Bulk_Eigen}) demonstrate the presence of few
eigenvalues outside the bulk region. Datasets of 1913-27 do not exhibit formation of bulk due to scarcity in number of data points.
Datasets of 1928-1952 depict separation of eigenvalues from bulk indicating existence of community structure (please refer main article
for elaboration).\\

\begin{suppfigure}[h]
\centering
\includegraphics[width=0.4\columnwidth]{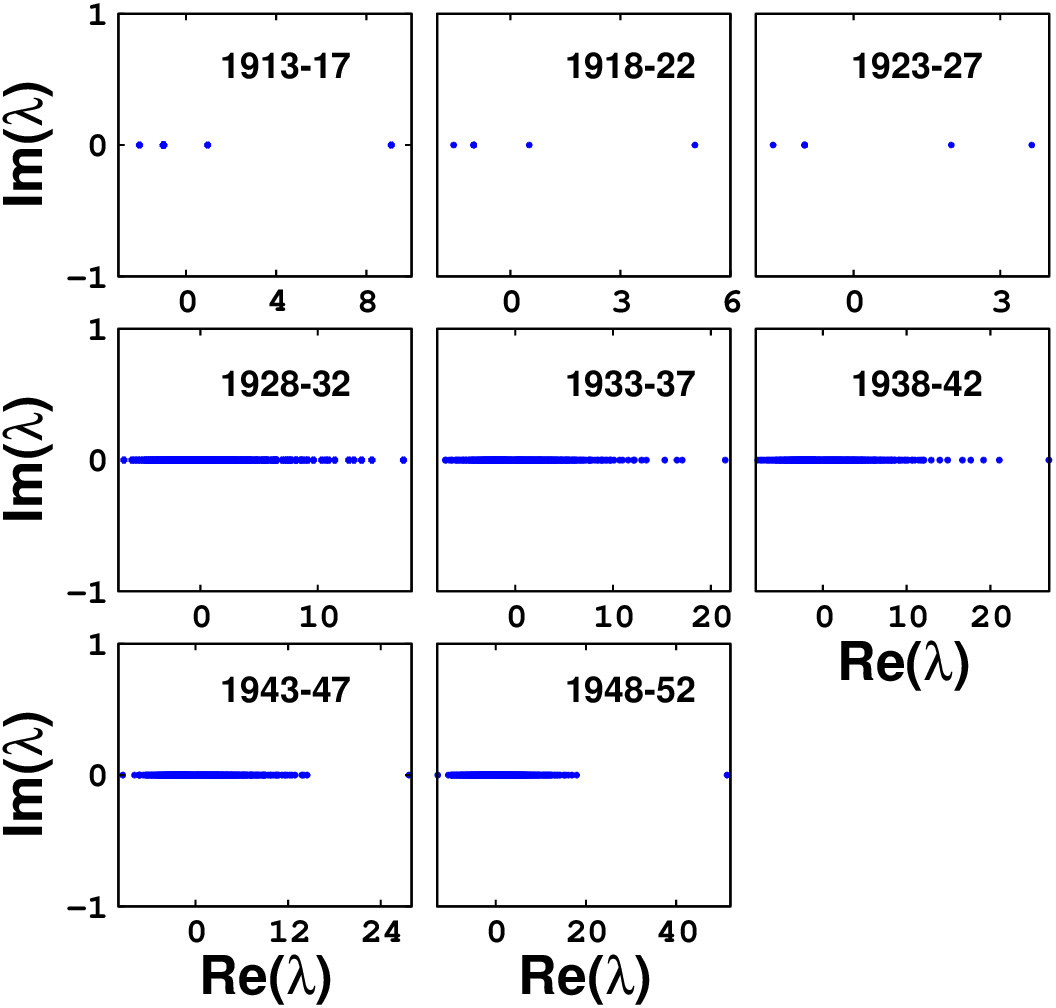}
\caption{Separation of lone eigenvalues from bulk of Bollywood datasets spanning between 1913-1952.}
\label{Bulk_Eigen}
\end{suppfigure}

\noindent {\textbf{1.3.1 Nearest neighbor spacing distribution (NNSD)}} \\

%\subsubsection{Nearest neighbor spacing distribution (NNSD)}
\noindent SI Fig.~\ref{nnsd} depicts NNSD of Bollywood networks. Discussion on NNSD is provided in the main article.\\

\noindent {\textbf{1.3.2 $\Delta_3$ Statistics}} \\
%\subsubsection{$\Delta_3$ Statistics}

\noindent It can be seen from SI Fig.~\ref{Fig_Delta_3} 
that the statistics agrees very well with the RMT prediction for 
some length for certain sets, and for some sets they do not follow RMT prediction of GOE 
statistics at all. The range for which $\Delta_3(L)$ statistics follows RMT prediction can be 
interpreted as providing measure of randomness in underlying network \cite{Jalan_2009}.
The length of the spectra which follow RMT prediction of GOE statistics is written in 
Table 1 of main article. In some of the sets namely 1953-57, 1958-62 and 1968-72 $\Delta_3(L)$ statistics does not 
follow RMT prediction at all. \\

\noindent {\large{\textbf{1.4 Net payoff}}} \\
%\subsection{Net payoff} 
%\vspace{0.1cm}

\noindent Net payoff is a measure originally borrowed from management 
which is modified and used as a predictive means for assessing success. 
PageRank algorithm has also been used to assign
ranks to nodes using a Markov chain based on the structure of the graph. 
This algorithm was used on Hollywood datasets to uncover the relative importance of a particular actor in the graph \cite{Vigna}.
The payoff defined here takes into account the essence of PageRank algorithm, alongwith other factors influencing 
the importance of a particular node.
Statistics supporting the net payoff of lead male actors
and female actors defined and discussed in the main article have been provided here in SI Tables 1-12. The 2003-07 span
defies the trend of positive correlation between overlaps of the male actors appearing in top
three consecutive positions of payoff list and their Filmfare nominations, 
where Amitabh Bachchan appears highest in the award nominees list. Here, it
would be noteworthy to mention that the legendary Padma Shri (1984), Padma Bhushan (2001), Amitabh
Bachchan (1969-2013), unlike all lead male actors of the yesteryear era, is the only one whose career never deteriorated.  
With 43 Filmfare nominations and being crowned as ``Superstar of the Millennium" in 2000 at the Filmfare Awards, he redeems to be
the superstar till date and is beyond all bounds. \\
\begin{suppfigure}[h!]
\vspace{-20pt}
        \centering
        \begin{subfigure}[b]{0.4\textwidth}
                \centering
            \includegraphics[width=\columnwidth]{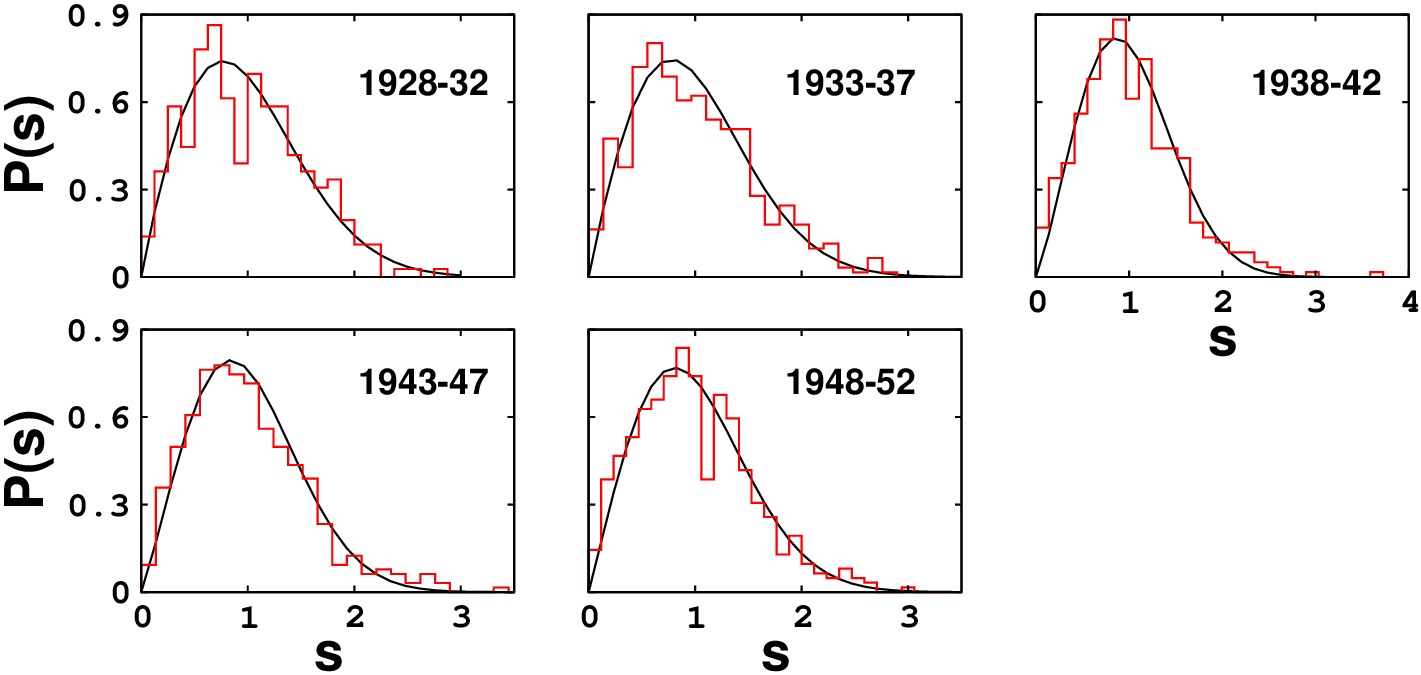}
        \end{subfigure}
        \quad 
        \begin{subfigure}[b]{0.4\textwidth}
                \centering
             \includegraphics[width=\columnwidth]{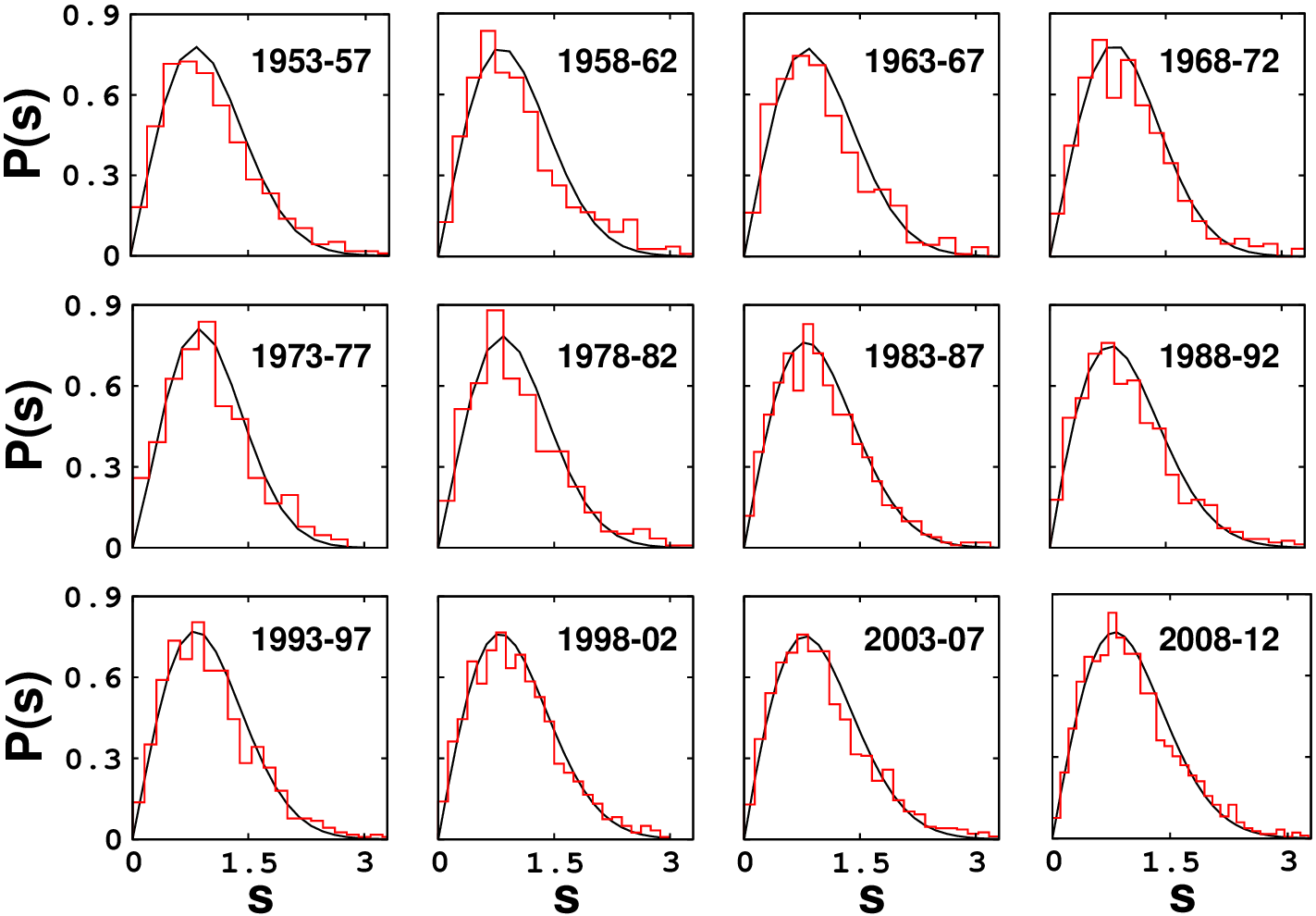}
        \end{subfigure}
	\vspace{-10pt}
        \caption{Nearest-neighbor spacing distribution $P(s)$ of the adjacency matrix of Bollywood networks. 
Histograms are numerical results and solid lines represent the NNSD of GOE.}
	\vspace{-5pt}
	\label{nnsd}
\end{suppfigure}

\noindent Lead female actors appearing in top five positions of net payoff list have been observed to bag the top 
three positions in terms of Filmfare award nominations (manually selected) which is very precise in the recent dataset 
where top five of net payoff correspond to top four nominated lead female actors, except for Katrina Kaif, who does not
have any Filmfare award nomination in 2003-2007 span still appearing at the 4th position in the top five (SI Table 7). 
She has been one of the most popular female actors in Bollywood since 2007, net payoff seems to be predictive of her success. \\
\begin{suppfigure}
        \centering
        \begin{subfigure}[b]{0.4\textwidth}
                \centering
            \includegraphics[width=\columnwidth]{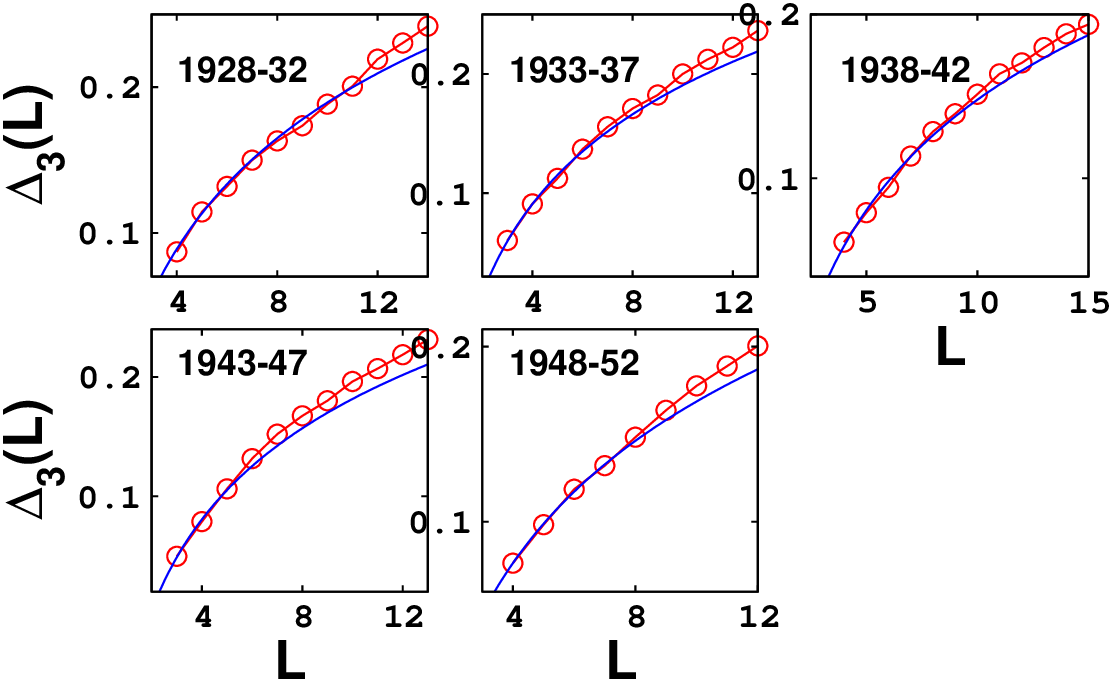}
        \end{subfigure}
        \quad 
        \begin{subfigure}[b]{0.4\textwidth}
                \centering
             \includegraphics[width=\columnwidth]{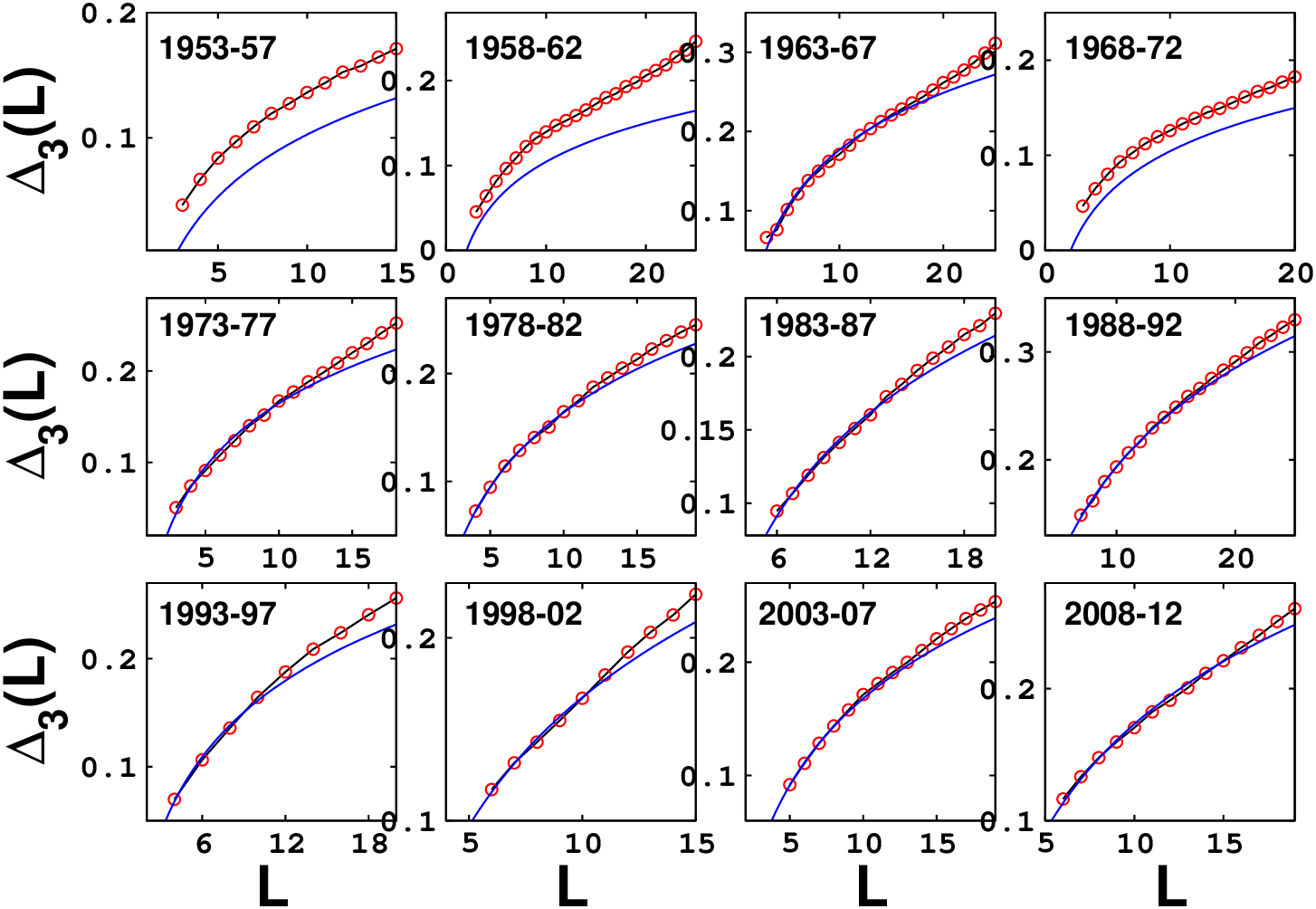}
        \end{subfigure}
	\vspace{-5pt}
        \caption{$\Delta_3(L)$ statistics of Bollywood networks. The solid line represents the GOE prediction, $\Delta_3(L)$ statistics 
follows the
RMT prediction up to length L.}
	\vspace{-5pt}
	\label{Fig_Delta_3}
\end{suppfigure}

\begin{supptable}[!htb]
    \caption{List of male actors holding top 10 positions in net payoff list of (a) and (b) datasets. Awards correspond to their award 
nominations in Filmfare in that particular span.}
    \begin{subtable}{.5\linewidth}
      \centering
        \caption{1953-1957}
\begin{tabular}{|p {3 cm}|p {1 cm}|l|}	\hline
Actors & $k$ 		&  Award(s) \\ \hline
Ashok Kumar	&	115	&	-	\\ \hline	
Balraj Sahni	&	115	&	-	\\ \hline
Raj Kapoor	&	78	&	1	\\ \hline
Dilip Kumar	&	115	&	3	\\ \hline
Shammi Kapoor	&	107	&	-	\\ \hline
Dev Anand	&	84	&	-	\\ \hline
Kishore Kumar	&	120	&	-	\\ \hline
Ajit		&	113	&	-	\\ \hline
Pradeep Kumar	&	114	&	-	\\ \hline
Mahipal		&	85	&	-	\\ \hline
        \end{tabular}
    \end{subtable}%
    \begin{subtable}{.5\linewidth}
      \centering
       \caption{1958-1962}
\begin{tabular}{|p {3 cm}|p {1 cm}|l|}	\hline
List of Actors & $k$ &  Award(s) \\ \hline
Ashok Kumar		&	156	&	-	\\ \hline	
Dev Anand		&	115	&	3	\\ \hline
Sunil Dutt		&	87	&	-	\\ \hline
Dharmendra		&	61	&	-	\\ \hline
Shammi Kapoor		&	114	&	-	\\ \hline
Manoj Kumar		&	73	&	-	\\ \hline
Rajendra Kumar		&	113	&	-	\\ \hline
Shashi Kapoor		&	48	&	-	\\ \hline
Pradeep Kumar		&	93	&	-	\\ \hline
Kishore Kumar		&	97	&	-	\\ \hline
        \end{tabular}
    \end{subtable} 
\end{supptable}

\begin{supptable}[!htb]
    \caption{List of male actors holding top 10 positions in net payoff list of (a) and (b) datasets. Awards correspond to their award 
nominations in Filmfare in that particular span.}
    \begin{subtable}{.5\linewidth}
      \centering
        \caption{1963-67}
\begin{tabular}{|p {3 cm}|p {1 cm}|l|}	\hline
List of Actors & k &  Award(s) \\ \hline
Dharmendra		&	191	&	2	\\ \hline	
Ashok Kumar		&	160	&	3	\\ \hline
Manoj Kumar		&	107	&	-	\\ \hline
Biswajeet		&	115	&	-	\\ \hline
Shashi Kapoor		&	103	&	-	\\ \hline
Dev Anand		&	88	&	1	\\ \hline
Sunil Dutt		&	110	&	2	\\ \hline
Sanjeev Kumar		&	61	&	-	\\ \hline
Dara Singh Randhawa	&	63	&	-	\\ \hline
Rajendra Kumar	&	72	&	4	\\ \hline
        \end{tabular}
    \end{subtable}%
    \begin{subtable}{.5\linewidth}
      \centering
       \caption{1968-72}
\begin{tabular}{|p {3 cm}|p {1 cm}|l|}	\hline
List of Actors & $k$ &  Award(s) \\ \hline
Amitabh Bachchan	&	178	&	1	\\ \hline	
Sanjeev Kumar		&	247	&	2	\\ \hline
Rajesh Khanna		&	234	&	5	\\ \hline
Vinod Khanna		&	179	&	-	\\ \hline
Shatrughan Sinha	&	204	&	1	\\ \hline
Dharmendra		&	221	&	1	\\ \hline
Jeetendra		&	204	&	-	\\ \hline
Shashi Kapoor		&	115	&	-	\\ \hline
Dara Singh Randhawa	&	156	&	-	\\ \hline
Dev Anand		&	119	&	-	\\ \hline
        \end{tabular}
    \end{subtable} 
\end{supptable}

\begin{supptable}[!htb]
    \caption{List of male actors holding top 10 positions in net payoff list of (a) and (b) datasets. Awards correspond to their award 
nominations in Filmfare in that particular span.}
    \begin{subtable}{.5\linewidth}
      \centering
        \caption{1973-77}
\begin{tabular}{|p {3 cm}|p {1 cm}|l|}	\hline
List of Actors & $k$ &  Award(s) \\ \hline
Rajesh Khanna		&	190	&	6	\\ \hline	
Sanjeev Kumar		&	234	&	5	\\ \hline
Dharmendra		&	258	&	2	\\ \hline
Amitabh Bachchan	&	299	&	4	\\ \hline
Shashi Kapoor		&	195	&	2	\\ \hline
Shatrughan Sinha	&	198	&	1	\\ \hline
Vinod Khanna		&	191	&	2	\\ \hline
Ashok Kumar		&	191	&	2	\\ \hline
Vinod Mehra		&	178	&	-	\\ \hline
Jeetendra		&	152	&	-	\\ \hline
        \end{tabular}
    \end{subtable}%
    \begin{subtable}{.5\linewidth}
      \centering
       \caption{1978-82}
\begin{tabular}{|p {3 cm}|p {1 cm}|l|}	\hline
List of Actors & $k$ &  Award(s) \\ \hline
Naseruddin Shah	&	117	&	3	\\ \hline	
Amitabh Bachchan	&	212	&	10	\\ \hline
Dharmendra		&	181	&	-	\\ \hline
Shashi Kapoor		&	226	&	-	\\ \hline
Rajesh Khanna		&	184	&	3	\\ \hline
Jeetendra		&	195	&	-	\\ \hline
Raj Babbar		&	179	&	1	\\ \hline
Sanjeev Kumar		&	221	&	5	\\ \hline
Shatrughan Sinha	&	185	&	2	\\ \hline
Om Puri		&	83	&	1	\\ \hline
        \end{tabular}
    \end{subtable} 
\end{supptable}

\begin{supptable}[!htb]
    \caption{List of male actors holding top 10 positions in net payoff list of (a) and (b) datasets. Awards correspond to their award 
nominations in Filmfare in that particular span.}
    \begin{subtable}{.5\linewidth}
      \centering
       \caption{1983-87}
\begin{tabular}{|p {3 cm}|p {1 cm}|l|}	\hline
List of Actors & $k$ &  Award(s) \\ \hline
Naseruddin Shah	&	218	&	5	\\ \hline	
Javed Khan		&	198	&	-	\\ \hline
Amitabh Bachchan	&	217	&	4	\\ \hline
Dharmendra		&	199	&	1	\\ \hline
Anil Kapoor		&	195	&	2	\\ \hline
Om Puri		&	207	&	1	\\ \hline
Suresh Oberoi		&	246	&	1	\\ \hline
Mithun Chakraborty	&	233	&	-	\\ \hline
Jackie Shroff		&	168	&	-	\\ \hline
Raj Babbar		&	278	&	2	\\ \hline
        \end{tabular}
    \end{subtable}%
    \begin{subtable}{.5\linewidth}
      \centering
      \caption{1988-92}
\begin{tabular}{|p {3 cm}|p {1 cm}|l|}	\hline
List of Actors & $k$ &  Award(s) \\ \hline
Mithun Chakraborty	&	302	&	1	\\ \hline	
Jackie Shroff		&	220	&	1	\\ \hline
Govinda		&	251	&	-	\\ \hline
Anil Kapoor		&	225	&	3	\\ \hline
Sanjay Dutt		&	249	&	1	\\ \hline
Jeetendra		&	196	&	-	\\ \hline
Rishi Kapoor		&	197	&	1	\\ \hline
Dharmendra		&	269	&	-	\\ \hline
Sunny Deol		&	155	&	1	\\ \hline
Akshay Kumar		&	83	&	-	\\ \hline
        \end{tabular}
    \end{subtable} 
\end{supptable}

\begin{supptable}[!htb]
    \caption{List of male actors holding top 10 positions in net payoff list of (a) and (b) datasets. Awards correspond to their award 
nominations in Filmfare in that particular span.}
    \begin{subtable}{.5\linewidth}
      \centering
      \caption{1993-97 }
\begin{tabular}{|p {3 cm}|p {1 cm}|l|}	\hline
List of Actors & $k$ &  Award(s) \\ \hline
Shahrukh Khan		&	225	&	7	\\ \hline	
Raza Murad		&	296	&	-	\\ \hline
Jackie Shroff		&	236	&	5	\\ \hline
Sanjay Dutt		&	162	&	1	\\ \hline
Kiran Kumar		&	324	&	1	\\ \hline
Suniel Shetty		&	167	&	1	\\ \hline
Naseruddin Shah	&	161	&	4	\\ \hline
Govinda		&	186	&	4	\\ \hline
Mithun Chakraborty	&	205	&	1	\\ \hline
Akshay Kumar		&	235	&	1	\\ \hline
        \end{tabular}
    \end{subtable}%
    \begin{subtable}{.5\linewidth}
      \centering
      \caption{1998-02}
\begin{tabular}{|p {3 cm}|p {1 cm}|l|}	\hline
List of Actors & $k$ &  Award(s) \\ \hline
Shahrukh Khan		&	291	&	8	\\ \hline	
Jackie Shroff		&	398	&	2	\\ \hline
Om Puri		&	286	&	3	\\ \hline
Sanjay Dutt		&	304	&	2	\\ \hline
Ajay Devgn		&	249	&	3	\\ \hline
Salman Khan		&	199	&	4	\\ \hline
Suniel Shetty		&	246	&	3	\\ \hline
Govinda		&	208	&	7	\\ \hline
Akshay Kumar		&	159	&	2	\\ \hline
Mithun Chakraborty	&	173	&	-	\\ \hline
        \end{tabular}
    \end{subtable} 
\end{supptable}

\begin{supptable}[!t]
\begin{center}
\caption{List of male actors holding top 10 positions in net payoff list of 2003-07 datasets. Awards correspond to their award nominations in Filmfare in that particular span.}
\begin{tabular}{|p {3 cm}|p {1 cm}|l|}	\hline
List of Actors & $k$ &  Award(s) \\ \hline
Salman Khan		&	261	&	3	\\ \hline	
Irrfan Khan		&	201	&	1	\\ \hline
Jackie Shroff		&	206	&	1	\\ \hline
Ajay Devgn		&	228	&	6	\\ \hline
Milind Gunaji		&	230	&	-	\\ \hline
Akshay Kumar		&	326	&	4	\\ \hline
Shahrukh Khan		&	246	&	9	\\ \hline
Shakti Kapoor		&	315	&	-	\\ \hline
Kay Kay Menon		&	216	&	1	\\ \hline
Sanjay Dutt		&	322	&	4	\\ \hline
\end{tabular}
\end{center}
\label{payoff_03}
\end{supptable}

\begin{supptable}[!t]
\begin{center}
\caption{List of female actors in descending order of their net payoffs in 2003-07 span who are manually 
selected based on their popularity, Filmfare award nominations, 
income {\it www.filmfare.com}. Award(s) correspond to their award nominations in 
Filmfare in that particular span.}
\begin{tabular}{|p {4 cm} |p {3.5 cm}|l|}	\hline
Name & Net payoff & Award(s)  \\ \hline
Kareena Kapoor	&	0.49	& 4	\\  \hline 
Priyanka Chopra	&	0.46	& 4	\\  \hline 
Rani Mukerji	&	0.44	& 10 	\\  \hline 
Katrina Kaif	&	0.39	& -	\\  \hline 
Bipasha Basu	&	0.37    & 4	\\  \hline 
\end{tabular}
\end{center}
\label{payoff_03_actress}
\end{supptable}

\begin{supptable}[!htb]
    \caption{List of female actors in descending order of their net payoff list of (a) and (b) datasets who are manually 
selected based on their popularity, Filmfare award nominations, 
income {\it www.filmfare.com}. Award(s) correspond to their award 
nominations in Filmfare in that particular span.}
    \begin{subtable}{.5\linewidth}
      \centering
      \caption{1998-02 }
	\begin{tabular}{|p {3 cm}|p {1 cm}|l|}	\hline
Name 			& 	Net payoff		& Award(s)  \\ \hline
Sridevi			&	0.72	& 1	\\  \hline
Rani Mukerji		&	0.58	& 2	\\  \hline
Tabu			&	0.54	& 7	\\  \hline
Mahima Choudhary 	&	0.53  	& 4 	\\  \hline
Aishwarya Rai		&	0.51	& 4	\\  \hline
 \end{tabular}
    \end{subtable}
    \begin{subtable}{.5\linewidth}
      \centering
      \caption{1993-97}
\begin{tabular}{|p {3 cm}|p {1 cm}|l|}	\hline
Name & Net payoff & Award(s) \\ \hline
Manisha Koirala	&	0.51	&  5	\\ \hline 
Raveena Tandon	&	0.47	&  1	\\ \hline 
Tabu		&	0.43	&  3 	\\ \hline 
Juhi Chawla	&	0.42	&  1	\\ \hline 
Madhuri Dixit	&	0.40    &  6	\\ \hline 
 \end{tabular}
    \end{subtable} 
\end{supptable}

\begin{supptable}[!htb]
    \caption{List of female actors in descending order of their net payoff list of (a) and (b) datasets who are manually 
selected based on their popularity, Filmfare award nominations, 
income {\it www.filmfare.com}. Award(s) correspond to their award 
nominations in Filmfare in that particular span.}
    \begin{subtable}{.5\linewidth}
      \centering
      \caption{1988-92 }
	\begin{tabular}{|p {3 cm}|p {1 cm}|l|}	\hline
Name & Net payoff & Award(s)  \\ \hline
Madhuri Dixit	 &	0.51	 &  4   \\ \hline
Juhi Chawla	 &	0.43	 &   2    \\ \hline
Dimple Kapadia	 &	0.33	 &  1    \\ \hline
Shilpa Shirodkar &	0.29	&  -  \\ \hline 
Farha		 &	0.28	&  -  \\ \hline 
\end{tabular}
    \end{subtable}
    \begin{subtable}{.5\linewidth}
      \centering
      \caption{1983-87}
\begin{tabular}{|p {3 cm}|p {1 cm}|l|}	\hline
Name & Net payoff & Award(s)  \\ \hline
Sadhana		   &	0.56    &  -  \\ \hline 
Rekha		   &	0.48	 &2 \\ \hline 
Meenakshi Seshadri &	0.45    &  -  \\ \hline 
Hema Malini	   &	0.45    &  -  \\ \hline 
Sridevi		   &	0.44	&  1 \\ \hline 
\end{tabular}
    \end{subtable} 
\end{supptable}

\begin{supptable}[!htb]
    \caption{List of female actors in descending order of their net payoff list of (a) and (b) datasets who are manually 
selected based on their popularity, Filmfare award nominations, 
income {\it www.filmfare.com}. Award(s) correspond to their award 
nominations in Filmfare in that particular span.}
    \begin{subtable}{.5\linewidth}
      \centering
      \caption{1978-82 }
	\begin{tabular}{|p {3 cm}|p {1 cm}|l|}	\hline
Name & Net payoff & Award(s)  \\ \hline
Rekha		&	0.67	 &   5 \\ \hline
Sarika	&	0.63	 &   1 \\ \hline
Hema Malini	&	0.63   &   3  \\ \hline
Parveen Babi	&	0.54   &  -   \\ \hline
Shabana Azmi	&	0.54	 &  2 \\ \hline
\end{tabular}
    \end{subtable}%
    \begin{subtable}{.5\linewidth}
      \centering
      \caption{1973-77 }
\begin{tabular}{|p {3 cm}|p {1 cm}|l|}	\hline
Name & Net payoff & Award(s)  \\ \hline
Rekha	        &	0.66	& -  \\  \hline	
Hema Malini	&	0.62	 & 6 \\  \hline	
Reena Roy	&	0.50	 & 1 \\  \hline	
Parveen Babi	&	0.49	& -  \\  \hline
Zeenat Aman	&	0.46	 & 1 \\  \hline	
\end{tabular}
    \end{subtable} 
\end{supptable}

\begin{supptable}[!htb]
    \caption{List of female actors in descending order of their net payoff list of (a) and (b) datasets who are manually 
selected based on their popularity, Filmfare award nominations, 
income {\it www.filmfare.com}. Award(s) correspond to their award 
nominations in Filmfare in that particular span.}
    \begin{subtable}{.5\linewidth}
      \centering
      \caption{1968-72}
\begin{tabular}{|p {3 cm} |p {1 cm}|l|}	\hline
Name & Net payoff & Award(s)  \\ \hline
Sulochana	&	0.78	&  -\\   \hline
Mumtaz		&	0.64	 & 3 \\   \hline 
Hema Malini	&	0.60	 & -    \\   \hline 
Jaya Bachchan	&	0.48    &  2\\   \hline  
Rekha		&	0.44  &  -\\   \hline 
\end{tabular}
    \end{subtable}
    \begin{subtable}{.5\linewidth}
      \centering
\caption{1963-67}
\begin{tabular}{|p {3 cm} |p {1 cm}|l|}	\hline
Name & Net payoff & Award(s)  \\ \hline
Sulochana	&	0.76   &  -\\   \hline
Mumtaz		&	0.74   &  -\\   \hline			
Mala Sinha	&	0.48	&  3\\   \hline
Meena Kumari	&	0.38	&  6\\   \hline
Tanuja  	&	0.38	&  -\\   \hline
\end{tabular}
    \end{subtable} 
\end{supptable}

\begin{supptable}[t]
    \caption{List of female actors in descending order of their net payoff list of (a) and (b) datasets who are manually 
selected based on their popularity, Filmfare award nominations, 
income {\it www.filmfare.com}. Award(s) correspond to their award 
nominations in Filmfare in that particular span.}
    \begin{subtable}{.5\linewidth}
      \centering
\caption{1958-62}
\begin{tabular}{|p {3 cm} |p {1 cm}|l|}	\hline
Name & Net payoff & Award(s)  \\ \hline
Sulochana	&	0.55	&  - \\   \hline
Mala Sinha	&	0.50	& 1\\   \hline
Mumtaz		&	0.46	&  - \\   \hline
Meena Kumari	&	0.40	& 1\\   \hline
Vyjayantimala	&	0.38    & 2\\   \hline
\end{tabular}
    \end{subtable}
    \begin{subtable}{.5\linewidth}
      \centering
\caption{1953-57}
\begin{tabular}{|p {3 cm} |p {1 cm}|l|}	\hline
Name & Net payoff & Award(s)  \\ \hline
Shyama	        &	0.72	& -\\   \hline
Meena Kumari	&	0.61	& 2\\   \hline
Sulochana	&	0.58  & -\\   \hline
Vyjayantimala	&	0.51  & 1\\   \hline
\end{tabular}
    \end{subtable} 
\end{supptable}


\begin{thebibliography}{99}
\bibitem{Barabasi_2002} Albert R, Barab\'asi AL (2002) Statistical mechanics of complex networks. Review of Modern Physics 74: 47-97.
\bibitem{Borgatti_2009} Borgatti SP, Mehra A, Brass DJ, Labianca G (2009) Network analysis in the social sciences. Science 323: 892-895.
\bibitem{Wasserman} Carrington PJ, Scott J, Wasserman S (2005) Models and methods in social network analysis.
Cambridge University Press, New York, 1st edition, 1-44 pp.
\bibitem{Barabasi_2010} Song C, Qu Z, Blumm N, Barab\'asi AL (2010) Limits of predictability in human mobility. Science 327: 1018-1021.
\bibitem{Focus_2010} KPMG India, Confederation of Indian Industry (2005) A CII-KPMG Report:
Indian Entertainment Industry Focus 2010-Dreams to Reality. KPMG India
and Confederation of Indian Industry.
\bibitem{Bose} Bose M (2007) Bollywood: A History, 1st edition. New Delhi: Rakmo Press. pp. 37–362.
\bibitem{FICCI_2011} KPMG India, Federation of Indian Chambers of Commerce and Industry(2011) FICCI-KPMG Indian Media and Entertainment Industry Report: Hitting the
High Notes. KPMG India and Federation of Indian Chambers of Commerce
and Industry.
\bibitem{Rev_Social_2006} Martino F, Spoto A (2006) Social network analysis: A brief theoretical review and further perspectives in the study of information technology. PsychNology Journal 4: 53-86.
\bibitem{Hollywood} Cattani G, Ferriani S (2006) A core/periphery perspective on individual creative performance:
Social networks and cinematic achievements in the hollywood film industry. Organization Science 4: 53-86.
\bibitem{Boldi_Vigna} Boldi P, Rosa M, Vigna S (2011) Robustness of social networks: Comparative results based on
distance distributions. Social Informatics 6984: 8-21.
\bibitem{FICCI_2012} KPMG India, Federation of Indian Chambers of Commerce and Industry(2012)
FICCI-KPMG Indian Media and Entertainment Industry Report: Digital Dawn
The metamorphosis begins. KPMG India and Federation of Indian Chambers
of Commerce and Industry.
\bibitem{Mehta_1991} Mehta ML (1991) Random Matrices, 2nd edition. New York: Academic Press.
\bibitem{handbook_rmt} Akemann G, Baik J, Francesco PD (2011) The Oxford Handbook of Random
Matrix Theory, 1st edition. Oxford: Oxford University Press.
\bibitem{SJ_2012} Jalan S, Ung CY, Bhojwani J, Li B, Zhang L, et al. (2012) Spectral analysis of gene co-expression
network of zebrafish. Europhysics Letters 99: e48004(1-6).
\bibitem{SJ_2007b} Jalan S, Bandyopadhyay JN (2007) Random matrix analysis of complex networks. Physical Review
E 76: e046107(1-7).
\bibitem{SJ_2009} Jalan S, Bandyopadhyay JN (2009) Randomness of random networks: A random matrix analysis.
Europhysics Letters 87: e48010(1-5).
\bibitem{Guhr} Guhr T, M-Groeling A, Weidenm$\ddot{u}$ller HA (1998) Random-matrix theories in quantum physics: common concepts. Physics Reports 299: 189-425.
\bibitem{URL} Bollywoodhungama website. Available: http://akm-www.bollywoodhungama.com/movies/list/sort/Rel
eased be listing/page/ and http://akm-www.bollywoodhungama.com/movies/list/sort/Released in 2012/
char/ALL/type/listing/page/. Accessed 2013 August 10.
\bibitem{filmfare} Filmfare website. Available: http://www.filmfare.com. Accessed 2013 August 10.
\bibitem{Newman_2003} Newman MEJ (2003) The structure and function of complex networks. SIAM Review 45: 167-256.
\bibitem{Jackson_1996} Jackson MO, Wolinsky A (1996) A strategic model of social and economic networks. Journal of
Economic Theory 71: 44-74.
\bibitem{A_Watts_2001} Watts A (2001) A dynamic model of network formation. Games and Economic Behavior 34: 331-
341.
\bibitem{Brody_1973} Brody TA (1973) Statistical measure for repulsion of energy-levels. Lett Nuovo Cimento 7: 482-
484.
\bibitem{task_force} University of Chicago (2006) Task Force Report: Economic reforms in India. Chicago, IL: University of Chicago.
\bibitem{Rita} Ray R (2012) Wither slumdog millionaire: India’s liberalization and development themes in bolly-
wood films. 17th International Business Research Conference, Toronto, Canada.
\bibitem{Pacheco} Pacheco JM, Santos FC, Chalub FACC (2006) Stern-judging: A simple, successful norm which
promotes cooperation under indirect reciprocity. PLoS Computational Biology 2: 1634-1638.
\bibitem{Lagoons} S\'uarez YR, J\'unior MP, Catella AC (2004) Factors regulating diversity and
abundance of fish communities in pantanal lagoons, brazil. Fisheries
Management and Ecology 11: 45–50.
\bibitem{SI} Supporting Information.
\bibitem{Aguiar_2005} de Aguiar MAM, Bar-Yam Y (2005) Spectral analysis and the dynamic response
of complex networks. Physical Review E 71: e016106(1–5).
\bibitem{book_spectra_graph} Mieghem PV (2011) Graph Spectra for Complex Networks, 1st edition. New
York: Cambridge University Press. pp. 11–345.
\bibitem{Dorogovtsev} Dorogovtsev SN, Goltsev AV, Mendes JFF, Samukhin AN (2003) Spectra of
complex networks. Physical Review E 68: e046109(1–10).
\bibitem{Chauhan_2009} Chauhan S, Girvan M, Ott E (2009) Spectral properties of networks with
community structure. Physical Review E 80: e056114(1–10).
\bibitem{nonlinear} Gammaitoni L, Hanggi P, Jung P, Marchesoni F (1998) Stochastic resonance.
Review of Modern Physics 70: 223–287.
\bibitem{Economic_crisis} Research Unit (LARRDIS), Rajya Sabha Secretariat, New Delhi (2009) Global
economic crisis and its impact on India. New Delhi, India.
\bibitem{gender_bias} Das D, Pathak M (2012) Gender equality: A core concept of socio-economic
development in india. Asian Journal of Social Sciences and Humanities 1: 257–
264.
\bibitem{book_gender} Kristof ND, WuDunn S (2009) Half the Sky: Turning Oppression into
Opportunity for Women Worldwide, 1st edition. New York: Vintage Publishing.
pp. 1–294.
\bibitem{Social_collaboration} Guimera R, Uzzi B, Spiro J, Amaral LAN (2005) Team assembly mechanisms
determine collaboration network structure and team performance. Science 308:
697–702.
\bibitem{industries} Tichy NM, Tushman ML, Fombrun C (1979) Social network analysis for
organizations. Academy of Management Review 4: 507–519.
\end{thebibliography}

\clearpage

\begin{thebibliography}{99}
\bibitem{Newman_2001} Newman M, E, J (2001) Scientific collaboration networks: I. Network construction and fundamental results. {\it Phys Rev E} 64: e016131(1-8).
\bibitem{Watts_1998} Watts D, J, Strogatz S, H (1998) Collective dynamics of
`small-world’ networks. {\it Nature} 393: 440-442.
\bibitem{Newman_2001b} Newman M, E, J, Strogatz S, H, Watts D, J (2001) Random graphs with arbitrary degree distributions and their applications. {\it Phys Rev E} 64: e026118(1-17).
\bibitem{RDN} Neville J, Jensen D (2007) Relational Dependency Networks. {\it JMLR} 8: 653-692.
\bibitem{Boldi_2010} Boldi P, Rosa M, Santini M, Vigna S (2011) Layered label propagation: A multiresolution coordinate-free ordering for compressing social networks. {\it Proc of the 20th international conference on World Wide Web}, (Hyderabad, India), pp 587-596.
\bibitem{Vigna} Boldi P, Rosa M, Vigna S (2011) Robustness of Social Networks: Comparative Results Based on Distance
Distributions. {\it Soc Inform} 6984: 8-21.
\bibitem{Barabasi_2000} Albert R, Barab\'asi A-L (2000) Topology of Evolving Networks: Local Events and Universality. {\it Phys Rev Lett} 85(24): 5234-5237. 
\bibitem{Barabasi_2002} Albert R, Barab\'asi A-L (2002) Statistical mechanics of complex networks. {\it Rev Mod Phys} 74(1): 47-97.
\bibitem{Newman_2003} Newman M, E, J (2003) The structure and function of complex networks. {\it SIAM Rev} 45(2): 167-256.
\bibitem{random_graph} Erd\H{o}s P, R\'enyi A (1960) On the Evolution of Random Graphs. {\it Publications of the Mathematical Institute of the Hungarian Academy of Sciences} 5: 17-61.
\bibitem{Barabasi_1999} Barab\'asi A-L, Albert R (1999) Emergence of scaling in random networks. {\it Science} 286: 509-512.
\bibitem{Farkas_2003} Farkas I, J, Der\'enyi I, Barab\'asi A-L, Vicsek T (2001) Spectra of ‘‘real-world’’ graphs: Beyond the semicircle law. {\it Phys Rev E} 64: e026704(1-12).
\bibitem{Dorogovstev_2003} Dorogovstev S, N, Goltsev A, V, Mendes J, F, F, Samukhin A, N (2003) Spectra of complex networks. {\it Phys Rev E} 68: e046109(1-10).
\bibitem{Mehta_1991} Mehta M, L (1991) {\it Random Matrices} (Academic Press, New York), 2nd Ed, pp 2-667.
\bibitem{Brody_1973} Brody T, A (1973) Statistical measure for repulsion of energy-levels. {\it Lett Nuovo Cimento} 7(12): 482-484.
\bibitem{Jalan_2009} Jalan S, Bandyopadhyay J, N (2009) Randomness of random networks: A random matrix analysis. {\it Europhys Lett} 87: e48010(1-5). 
\end{thebibliography}
\end{document}